\documentclass[useAMS,usenatbib]{mn2e}

\usepackage{epsfig,color,pifont}

\usepackage{natbib}
\bibpunct{(}{)}{;}{a}{}{,}
\def\spose#1{\hbox to 0pt{#1\hss}}
\def\ltsimm{\mathrel{\spose{\lower 3pt\hbox{$\sim$}}
        \raise 2.0pt\hbox{$<$}}}
\def\gtsimm{\mathrel{\spose{\lower 3pt\hbox{$\sim$}}
        \raise 2.0pt\hbox{$>$}}}

\def\cm{{\rm\thinspace cm}}

\def\s{{\rm\thinspace s}}

\def\g{{\rm\thinspace g}}

\def\erg{{\rm\thinspace erg}}
\def\Hz{{\rm\thinspace Hz}}

\def\ster{{\rm\thinspace ster}}
\def\ergps{\hbox{${\rm\erg\s^{-1}\,}$}}

\def\pcm{\hbox{${\rm\cm^{-1}\,}$}}
\def\pcm2{\hbox{${\rm\cm^{-2}\,}$}}
\def\pcm3{\hbox{${\rm\cm^{-3}\,}$}}
\def\ergpscm3Hz{\hbox{${\rm\ergps\cm^{-3}\Hz^{-1}\,}$}}
\def\ergpscm3Hzster{\hbox{${\rm\ergps\cm^{-3}\Hz^{-1}\ster^{-1}\,}$}}
\def\gpcm3{\hbox{${\rm\g\cm^{-3}\,}$}}
\def\ergpcm2{\hbox{${\rm\erg\cm^{-2}\,}$}}
\def\ergpcm3{\hbox{${\rm\erg\cm^{-3}\,}$}}
\def\phpscm2{\hbox{${\rm photons\s^{-1}\cm^{-2}\,}$}}

\def\aap{{\rm A\&A}}
\def\apj{{\rm ApJ}}
\def\apjl{{\rm ApJL}}
\def\apjs{{\rm ApJS}}

\def\mnras{{\rm MNRAS}}

\def\araa{{\rm ARA\&A}}

\def\jcp{{\rm J.~Comput.~Phys}}

\def\actaa{{\rm AcA}}
\def\newa{{\rm NewA}}

\title [Global MHD disks I]{Global simulations of magnetorotational
  turbulence I: convergence and the quasi-steady state}

\author[E.~R.~Parkin \& G.~V.~Bicknell]
{E.~R.~Parkin\thanks{E-mail:parkin@mso.anu.edu.au} \&
  G.~V.~Bicknell\\Research School of Astronomy and Astrophysics,
  Australian National University, Canberra, ACT 2611, Australia}

\begin{document}

\date{Accepted ... Received ...; in original form ...}

\pagerange{\pageref{firstpage}--\pageref{lastpage}} \pubyear{2013}

\maketitle

\label{firstpage}

\begin{abstract}
  Magnetorotational turbulence provides a viable mechanism for angular
  momentum transport in accretion disks. We present global, three
  dimensional (3D), magnetohydrodynamic accretion disk simulations
  that investigate the dependence of the turbulent stresses on
  resolution. Convergence in the time-and-volume-averaged
  stress-to-gas-pressure ratio, $\overline{\langle\alpha_{\rm
      P}\rangle}$, at a value of $\sim0.04$, is found for a model with
  radial, vertical, and azimuthal resolution of 12-51, 27, and 12.5
  cells per scale-height (the simulation mesh is such that cells per
  scale-height varies in the radial direction). The gas pressure
  dependence of the quasi-steady state stress level is also examined
  using models with different scaleheight-to-radius aspect ratio
  ($H/R$), revealing a weak dependence of
  $\overline{\langle\alpha_{\rm P}\rangle}$ on pressure.

  A control volume analysis is performed on the main body of the disk
  ($|z|<2H$) to examine the production and removal of magnetic
  energy. Maxwell stresses in combination with the mean disk rotation
  are mainly responsible for magnetic energy production, whereas
  turbulent dissipation (facilitated by numerical resistivity)
  predominantly removes magnetic energy from the disk. Re-casting the
  magnetic energy equation in terms of the power injected by Maxwell
  stresses on the boundaries of, and by Lorentz forces within, the
  control volume highlights the importance of the boundary conditions
  (of the control volume). The different convergence properties of
  shearing-box and global accretion disk simulations can be readily
  understood on the basis of choice of boundary conditions and the
  magnetic field configuration. Periodic boundary conditions restrict
  the establishment of large-scale gradients in the magnetic field,
  limiting the power that can be delivered to the disk by Lorentz
  forces and by stresses at the surfaces. The factor of three lower
  resolution required for convergence in $\langle\alpha_{\rm
    P}\rangle$ for our global disk models compared to stratified
  shearing-boxes is explained by this finding.
\end{abstract}




\section{Introduction}
\label{sec:intro}

For the astrophysically common process of mass accretion through a
disk to be effective, outward angular momentum transport must occur
\citep[][]{Shakura:1973,Pringle:1981}. In the past two decades it has
become clear that self-sustaining magnetized turbulence driven by the
magnetorotational instability (MRI) can play this role \citep{BH98}.

Due to the highly non-linear nature of magnetorotational turbulence,
numerical simulations have become a common tool in its study. These
simulations come in a number of flavours: unstratified shearing-boxes
(where the vertical component of gravity is
neglected)\citep{Hawley:1995, Fromang:2007a, Fromang:2007b,
  Lesur:2007, Lesur:2011, Lesaffre:2009, Latter:2009,
  Heinemann_Papaloizou:2009, Simon:2009, Guan:2009, Bodo:2011,
  Kapyla:2011, Latter:2012}, stratified shearing-boxes
\citep{Brandenburg:1995, Stone:1996, Miller:2000, Fleming:2000,
  Brandenburg:2005, Johansen:2009, Gressel:2010, Shi:2010, Davis:2010,
  Simon:2011, Guan:2011, Oishi:2011, Simon:2012}, unstratified global
models \citep{Hawley:2001, Armitage:2001, Nelson:2010, Sorathia:2012},
and stratified global models \citep{Hawley:2000, Hawley_Krolik:2001,
  Arlt:2001, Fromang:2006, Fromang:2009, Beckwith:2008, Lyra:2008,
  Sorathia:2010, O'Neill:2011, Flock:2011, Flock:2012, Noble:2010,
  Beckwith:2011, Hawley:2011, McKinney:2012,
  Parkin:2013}. Shearing-box simulations focus on a local patch of an
accretion disk whereas global simulations have the potential to study
the entire radial (and vertical) extent of an accretion disk. Despite
these numerous different approaches to modelling accretion disk
turbulence, similarities exist in the magnetorotational turbulence
that they exhibit. In general, there is an initial phase where the MRI
develops and transient magnetic field amplification arises, following
which the growth of stresses subsides and the disk settles into a
quasi-steady state (QSS).

There have been mixed results from simulations as to what sets the QSS
stress level. The results of unstratified shearing-box simulations by
\cite{Fromang:2007a} \citep[see also-][]{Lesur:2007, Fromang:2007b,
  Simon:2009, Guan:2009, Fromang:2010, Kapyla:2011} show that
dissipation (i.e. resistivity and viscosity) dictates the QSS stress
level. When this dissipation is purely numerical in origin, increasing
the simulation resolution causes a reduction in the volume averaged
stress in zero-net flux, unstratified shearing-box
simulations. \cite{Fromang:2007a} argue that this occurs because
magnetorotational turbulence always drives energy to the smallest
resolved scale, thus removing energy from the larger (angular momentum
transporting) eddies. \cite{Sorathia:2012} have recently revisited
this issue using unstratified global disks, revealing a contrasting
result of converged stresses with increasing resolution. What then
sets the QSS stress level? \cite{Vishniac:2009} has argued that
stratification, if present, will affect the QSS stress, and it is
indeed found that including stratification facilitates convergence
\citep{Davis:2010,Shi:2010,Oishi:2011}. Furthermore, including a net
flux field in unstratified shearing-boxes enables convergence
\citep[e.g.][]{Simon:2009}. Considering the aforementioned results,
there is a clear indication that the choice of numerical setup and/or
magnetic field configuration play crucial roles.

In this work we take the logical next step and investigate convergence
in stratified global disk models, which is indeed found, but for lower
resolutions than in equivalent shearing-box simulations. A
complementary analysis of magnetic energy production leads us to
conclude that boundary conditions have a profound influence on the QSS
stress. The remainder of this paper is organised as follows: in
\S~\ref{sec:model} we describe the simulation setup and diagnostics
used in this investigation. In \S~\ref{sec:saturation} we examine the
dependence of the saturated turbulent state on simulation resolution
and disk scale-height. The results from the application of a
control-volume analysis to the simulations are presented in
\S~\ref{sec:control_volume}. We discuss our findings in the context of
a unified interpretation for magnetorotational turbulence in different
numerical setups in \S~\ref{sec:convergence} and close with conclusions
in \S~\ref{sec:conclusions}.

\section{The model}
\label{sec:model}

\subsection{Simulation code}
\label{subsec:hydromodel}

The time-dependent equations of ideal magnetohydrodynamics are solved
using the {\sevensize PLUTO} code \citep{Mignone:2007} in a 3D spherical
$(r,\theta,\phi)$ coordinate system. The relevant equations for mass,
momentum, energy conservation, and magnetic field induction are:
\begin{eqnarray}
\frac{\partial\rho}{\partial t} + \nabla \cdot [\rho {\bf v}] &  =  & 0, \\
\frac{\partial\rho{\bf v}}{\partial t} + \nabla\cdot[\rho{\bf vv} -
{\bf BB} + P{\bf I}] & = & -\rho \nabla\Phi,\\
\frac{\partial E}{\partial t} + \nabla\cdot[(E + P){\bf v} - ({\bf v \cdot B}){\bf B}] &
=& -\rho {\bf v}\cdot \nabla\Phi -\Lambda \label{eqn:energy}\\
\frac{\partial\bf B}{\partial t} & = & \nabla \times ({\bf v \times B}). \label{eqn:induction}
\end{eqnarray}
\noindent Here $E = \rho\epsilon + \frac{1}{2}\rho|{\bf v}|^{2} +
u_{\rm B}$ is the total energy, $\epsilon$ is the internal energy,
${\bf v}$ is the velocity, $\rho$ is the mass density, $P$ is the
pressure, and $u_{\rm B}=\frac{1}{2}|B|^2$ is the magnetic energy. We
use an ideal gas equation of state, $P = (\gamma - 1)\rho \epsilon$,
with an adiabatic index $\gamma = 5/3$. The adopted scalings for density,
velocity, temperature, and length are, respectively,
\begin{eqnarray}
  \rho_{\rm scale}&=&1.67\times10^{-7}~{\rm gm~s^{-1}}, \nonumber \\
  v_{\rm scale}&=&c, \nonumber \\
  T_{\rm scale}&=&\mu m c^{2}/k_{\rm B} = 6.5\times10^{12}~{\rm K},
  \nonumber \\
  l_{\rm scale}&=&1.48\times10^{13}~{\rm cm}, \nonumber 
\end{eqnarray}
\noindent where $c$ is the speed of light, and the value of $l_{\rm
  scale}$ corresponds to the gravitational radius of a $10^{8}~{\rm
  M_{\odot}}$ black hole.

The gravitational potential due to a central point mass situated at
the origin, $\Phi$, is modelled using a pseudo-Newtonian potential
\citep{Paczynsky:1980}:
\begin{equation}
\Phi = \frac{-1}{r - 2}.
\end{equation}
\noindent
Note that we take the gravitational radius (in scaled units), $r_{\rm
  g} = 1$. The Schwarzschild radius, $r_{\rm s}=2$ for a spherical
black hole and the innermost stable circular orbit (ISCO) lies at
$r=6$. The $\Lambda$ term on the RHS of Eq~(\ref{eqn:energy}) is an
ad-hoc cooling term used to keep the scale-height of the disk
approximately constant throughout the simulations; without any
explicit cooling in conjunction with an adiabatic equation of state,
dissipation of magnetic and kinetic energy leads to an increase in gas
pressure and, consequently, disk scale-height over time. The form of
$\Lambda$ used is identical to that of \cite{Parkin:2013}; further
details can be found in that paper\footnote{ Note that there is a
    typographical error in equation (3) of \cite{Parkin:2013} where
    $\rho\Lambda$ should read $\Lambda$.}.

The {\sevensize PLUTO} code was configured to use the five-wave HLLD
Riemann solver of \cite{Miyoshi:2005}, piece-wise parabolic
reconstruction \citep[PPM -][]{Colella:1984}, limiting during
reconstruction on characteristic variables \citep[e.g.][]{Rider:2007},
second-order Runge-Kutta time-stepping, and the upwind {\sevensize
  CONTACT} Constrained Transport scheme of \cite{Gardiner:2008} (to
maintain ${\bf \nabla \cdot B} = 0$) which includes transverse
corrections to interface states. This configuration was found to be
stable for linear MRI calculations by \cite{Flock:2011}.

The grid used for the simulations is uniform in the $r$ and $\phi$
directions and extends from $r=4-34$ and $\phi = 0 - \pi /2$. A graded
mesh is used in the $\theta$ direction which is uniform within $|z|
\leq 2H$ and stretched between $2H \le|z|\le 5H$, where $H$ is the
thermal disk scale-height. For our fiducial model, gbl-sr, there are a
total of 170 cells in the $\theta$ direction, of which 108 are
uniformly distributed within $|z| \le 2H$, and the remaining 62 cells
on the stretched sections between $2H\le |z|\le 5H$. Details of the
grid resolutions used in the simulations are provided in
Table~\ref{tab:model_list}. The adopted boundary conditions are
identical to those used in \cite{Parkin:2013}. Finally, floor density
and pressure values are used which scale linearly with radius and have
values at the outer edge of the grid of $10^{-4}$ and
$5\times10^{-9}$, respectively.

\subsection{Initial conditions}
\label{subsec:initialconditions}

The simulations start with an analytic equilibrium disk which is
isothermal in height ($T=T(R)$, where $T$ is the temperature) and
possesses a purely toroidal magnetic field. The derivation of the disk
equilibrium and a detailed description of the initial conditions can
be found in \cite{Parkin:2013}. In cylindrical coordinates ($R,z$),
the density distribution, in scaled units, is given by,
\begin{equation}
  \rho(R,z) = \rho(R,0) \exp\left(\frac{-\{\Phi(R,z) - \Phi(R,0)\}}{T(R)} \frac{\beta}{1 + \beta} \right), \label{eqn:rho}
\end{equation}
\noindent where the pressure, $P = \rho T$, and the ratio of
gas-to-magnetic pressure, $\beta = 2P/|B|^2 \equiv 2P/B_{\phi}^2$ is
initially set to 20 in all models. For the radial profiles $\rho(R,0)$
and $T(R)$ we use simple functions inspired by the \cite{Shakura:1973}
disk model, except with an additional truncation of the density
profile at a specified outer radius:
\begin{eqnarray}
  \rho(R,0) & = & \rho_{0} f(R,R_{0},R_{\rm out})
  \left(\frac{R}{R_{0}}\right)^{\epsilon}, \\
  T (R) & = & T_{\rm 0} \left(\frac{R}{R_{0}}\right)^{\eta}, \label{eqn:temp}
\end{eqnarray}
\noindent where $\rho_{0}$ sets the density scale, $R_{0}$ and $R_{\rm
  out}$ are the radius of the inner and outer disk edge, respectively,
$f(R,R_{0},R_{\rm out})$ is a tapering function \citep{Parkin:2013},
and $\epsilon$ and $\chi$ set the slope of the density and temperature
profiles, respectively. In all of the global simulations $R_{0}=7$,
$R_{\rm out}=30$, $\rho_{0}=10$, $\epsilon=-33/20$, and
$\eta=-9/10$. In \S~\ref{sec:saturation}, models with aspect ratios of
$H/R=0.05$ and 0.1 are considered. These ratios are achieved by
setting $T_{0}=4.5\times10^{-4}$ and $1.5\times10^{-3}$,
respectively. The rotational velocity of the disk is close to
Keplerian, with a minor modification due to the gas and magnetic
pressure gradients,
\begin{eqnarray}
  v_{\phi}^2(R,z) = v_{\phi}^2(R,0) + \{\Phi(R,z) -
  \Phi(R,0)\}\frac{R}{T}\frac{d T}{d R}, \label{eqn:vphi}
\end{eqnarray}
\noindent where,
\begin{eqnarray}
v_{\phi}^2(R,0) = R \frac{\partial
  \Phi (R,0)}{\partial R} + \frac{2 T}{\beta} + \nonumber \\
 \left(\frac{1 + \beta}{\beta} \right)\left(\frac{R
    T}{\rho(R,0)}\frac{\partial \rho(R,0)}{\partial R}
 + R \frac{d T}{d R}\right).   \label{eqn:vphi0}
\end{eqnarray}
The region outside of the disk is set to be an initially stationary,
spherically symmetric, hydrostatic atmosphere. The transition between
the disk and background atmosphere occurs where their total pressures
balance. To initiate the development of turbulence in the disk, a low
wavenumber, non-axisymmetric Fourier mode is excited in the poloidal
velocities with amplitude $0.1~c_{\rm s}$, where $c_{\rm s}$ is the
sound speed.

\subsection{Diagnostics}
\label{subsec:diagnostics}

A volume-averaged value (denoted by angled brackets) for a variable
$q$ is computed via,
\begin{equation}
  \langle q\rangle = \frac{\int q r^2 \sin \theta dr d\theta d\phi}{\int r^2 \sin
    \theta dr d\theta d\phi}.
\end{equation}
Similarly, azimuthal averages are denoted by square brackets,
\begin{equation}
[q] = \frac{\int q r \sin\theta d\phi}{ \int r \sin \theta d\phi}.
\end{equation}
\noindent Time averages receive an overbar, such that a volume and
time averaged quantity reads $\overline{\langle q\rangle}$. Throughout
this paper we concentrate on the region between $10<r<30$ and $\pi/2 -
\theta_{2H/R} <\theta< \pi/2 + \theta_{2H/R}$, where
$\theta_{2H/R}=\tan^{-1}(2\langle H/R\rangle)$ and $H/R = c_{\rm
  s}/v_{\phi}$ (where $c_{\rm s}$ is the sound speed). We define this
region as the ``disk body''.

The efficiency of angular momentum transport is typically quantified
from the total stress,
\begin{equation}
  W_{\rm ij} = G_{\rm ij} - M^{B}_{\rm ij}, 
\end{equation} 
where the Reynolds stress tensor,
\begin{equation}
  G_{\rm ij} = \rho \delta v_{\rm i}\delta v_{\rm j}, 
\end{equation} 
and the Maxwell stress tensor,
\begin{equation}
  M^{\rm B}_{\rm ij} = B_{\rm i}B_{\rm j} -  \delta_{\rm ij} u_{\rm B}. \label{eqn:maxstress}
\end{equation}
The largest contribution comes from the $R-\phi$ component of $W_{\rm
  ij}$ \citep{Brandenburg:1995, Hawley:1995, Stone:1996},
\begin{equation}
  W_{\rm R\phi} = \rho \delta v_{\rm R}\delta v_{\phi} - B_{\rm R}B_{\phi},
\end{equation}
where we have defined the perturbed flow velocity as\footnote{Using an
  azimuthally averaged velocity when calculating the perturbed
  velocity removes the influence of strong vertical and radial
  gradients \citep{Flock:2011}.}  $\delta v_{\rm i}=v_{\rm i}-[v_{\rm
  i}]$, with ${\rm i}={\rm R},\phi$. Normalising by the gas pressure
defines the $\alpha$-parameter,
\begin{equation}
  \langle \alpha_{\rm P}\rangle = \frac{\langle
    W_{R\phi}\rangle}{\langle P\rangle}.
\end{equation}
\noindent Furthermore, we calculate the $R-\phi$ component of the
Maxwell stress normalised by the magnetic pressure,
\begin{equation}
  \langle \alpha_{\rm M}\rangle = \frac{-\langle M^{\rm B}_{\rm
      R\phi}\rangle }{\langle u_{\rm B}\rangle } = \frac{\langle -2
    B_{\rm R}B_{\phi}\rangle }{\langle |B|^2\rangle }.
\end{equation}

We follow \cite{Noble:2010} and \cite{Hawley:2011} and utilize a
``quality factor'' to measure the ability of the simulations to
resolve the wavelength of the fastest growing MRI mode, $\lambda_{\rm
  MRI}$. Defining,
\begin{equation}
  \lambda_{\rm MRI-i} = \frac{2 \pi |v_{\rm A i}| r \sin \theta}{v_{\phi}},
\end{equation}
where $i=r,\theta,\phi$, and $v_{\rm A i}=B_{\rm i}/\sqrt{\rho}$ is
the Alfv{\' e}n speed, the ``quality factor'' is given by,
\begin{equation}
  Q_{\rm i}= \frac{\lambda_{\rm MRI-i}}{\Delta x_{\rm i}}, 
\end{equation}
where $\Delta x_{\rm i}$ is the cell spacing in direction $i$. The
``resolvability'' - the fraction of cells in the disk body that have
$Q>8$ \citep[e.g. ][]{Sorathia:2012} - is then defined as,
\begin{equation}
  N_{\rm i} = \frac{\Sigma C(Q_{\rm i}>8)}{\Sigma C} \label{eqn:resolvability}
\end{equation}
\noindent where $C$ represents a cell. 

\begin{table}
\begin{center}
  \caption[]{List of global simulations.} \label{tab:model_list}
\begin{tabular}{llllll}
\hline
  Model & $H/R$ & Resolution & $n_{\rm r}/H$ & $n_{\theta}/H$ & $n_{\phi}/H$ \\ 
 &  & ($n_{\rm r}$, $n_{\theta}$, $n_{\phi}$) & &  ($|z| < 2H$)  & \\
\hline
gbl-lr & 0.1 & 340,112,128 & 8.5-36 & 18 & 8 \\
gbl-sr & 0.1 & 512,170,196 & 12-51 & 27 & 12.5\\
gbl-hr &  0.1 & 768,256,256 & 18-77 & 37 & 16 \\
gbl-lr-la & 0.1 & 420,140,70 & 10.5-45 & 20 & 4.5 \\
gbl-hr-la & 0.1 & 768,256,128 & 18-77 & 37 & 8 \\
gbl-thin & 0.05 & 512,170,320 & 6-25 & 27 & 10 \\
\hline
\end{tabular}
\end{center}
\end{table}

\subsection{Fourier analysis}

The simulation data is Fourier transformed in spherical coordinates to
compute power spectra for different simulation variables. A detailed
description of the method used is given in Appendix~\ref{sec:fts}. In
brief, we define the Fourier transform of a function
$f(r,\theta,\phi)$ as,
\begin{eqnarray}
F({\bf k}) = F(k, \chi, \psi) = \int_0^{2 \pi} \int_0^\pi \int_0^\infty f(r,\theta,\phi) \, e^{i {\bf k \cdot x}} \times\nonumber\\
r^2 \sin \theta \, dr \, d \theta \, d \phi. \label{eqn:ft}
\end{eqnarray}
It then follows that the angle-averaged (in Fourier space) amplitude
spectrum,
\begin{equation}
\Pi(k) = \int_0^{2 \pi} \int_0^\pi F({\bf k}) F^*({\bf k}) \> \sin \chi \, d\chi \, d \psi, 
\end{equation}
where an asterisk ($^*$) indicates a complex conjugate. The total
power at a given wavenumber - the power spectrum - is then given by
$k^2 \Pi(k)$.To compute a power spectrum, we take 300 simulation
checkfiles (equally spaced over the last $15~P^{\rm orb}_{30}$, where
$P^{\rm orb}_{r}$ is the orbital period at a radius $r$) and compute
30 power-spectra, each time-averaged over $0.5~P^{\rm
  orb}_{30}$. These 30 realisations are then averaged-over to produce
the final power spectrum.

\begin{table*}
\begin{center}
  \caption[]{List of time averaged quantities from the global
    simulations. $\Delta t_{\rm av}$ (second column) is the time
    interval over which time averaging was
    performed.} \label{tab:global_models}
\begin{tabular}{lllllllllll}
  \hline
  Model & $\Delta t_{\rm av}$ & $\overline{N_{\rm r}}$ & $\overline{N_{\theta}}$ &
  $\overline{N_{\phi}}$ & $\overline{\langle \beta_{\rm r}\rangle}$ & $\overline{\langle\beta_{\theta}\rangle}$ &
  $\overline{\langle\beta_{\phi}\rangle}$ & $\overline{\langle\beta_{\rm d}\rangle}$ & $\overline{\langle\alpha_{\rm P}\rangle}$ & $\overline{\langle\alpha_{\rm M}\rangle}$ \\
  \hline
  gbl-lr & 12-31 & 0.57 & 0.27 & 0.67 & 131 & 395 & 14 & 12 & 
  0.043 & 0.39 \\ 
  gbl-sr & 12-31 & 0.69 & 0.43 & 0.75 & 128 & 361 & 17 & 14 & 
  0.040 & 0.42 \\
  gbl-hr & 12-31 & 0.78 & 0.56 & 0.80 & 123 & 332 & 18 & 15 & 
  0.039 & 0.42 \\
  gbl-lr-la & 18-31 & 0.31 & 0.07 & 0.25 & 655 & 2296 & 34 & 32 & 
  0.013 & 0.29 \\
  gbl-hr-la & 12-31 & 0.76 & 0.50 & 0.61 & 150 & 430 & 18 & 15 & 
  0.035 & 0.39 \\
  gbl-thin & 12-31 & 0.52 & 0.44 & 0.75 & 120 & 416 & 15 & 13 & 
  0.044 & 0.41 \\
  \hline
\end{tabular}
\end{center}
\end{table*}

\subsection{Summary of models}

In Table~\ref{tab:model_list} we list six simulations aimed at
investigating the following points:
\begin{itemize}
\item {\it Convergence with resolution:} Models gbl-lr, gbl-sr, and
  gbl-hr are low, standard, and high resolution variants,
  respectively, with identical cell aspect ratio and $H/R=0.1$. 
\item {\it Importance of azimuthal resolution:} Model gbl-lr-la
  (gbl-hr-la) is identical to gbl-lr (gbl-hr) with the exception of a
  lower azimuthal resolution (denoted by the affix ``-la'').
\item {\it Scale-height dependence:} Models gbl-sr and gbl-thin have
  disk scale-heights of $H/R=0.1$ and 0.05, respectively. These models
  feature an identical number of cells per scale-height in the vertical
  and azimuthal directions.
\end{itemize}

\begin{figure}
  \begin{center}
    \begin{tabular}{c}
\resizebox{75mm}{!}{\includegraphics{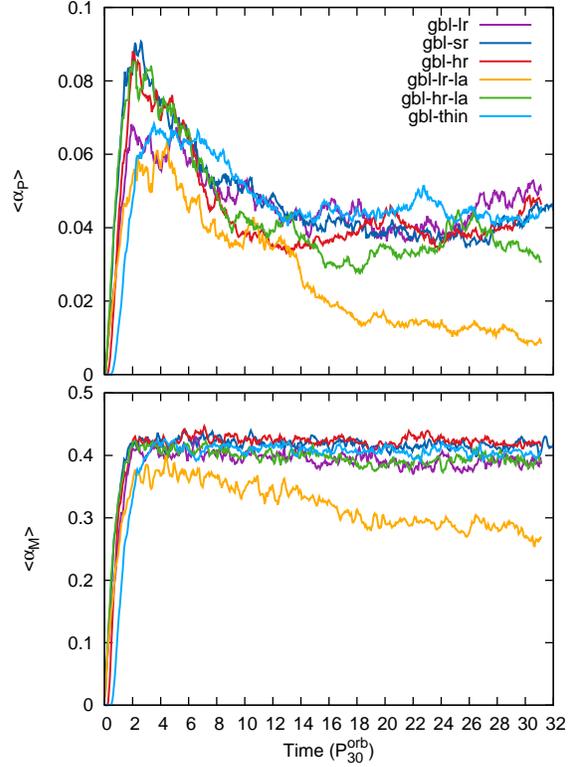}} \\
   \end{tabular}
   \caption{The time evolution of $\langle\alpha_{\rm P}\rangle$ (upper) and
     $\langle\alpha_{\rm M}\rangle$ (lower) in the global models, where time is in
     units of the orbital period at a radius of $r=30$, $P^{\rm
       orb}_{30}$. (For comparison, $P^{\rm orb}_{30}=11.6~P^{\rm
       orb}_{7}$, therefore roughly 370 inner disk orbits are
     covered.) Details pertaining to the models are listed in
     Table~\ref{tab:model_list} and corresponding time averaged
     results are given in Table~\ref{tab:global_models}.}
    \label{fig:alpha}
  \end{center}
\end{figure}

\begin{figure}
  \begin{center}
    \begin{tabular}{c}
    \resizebox{75mm}{!}{\includegraphics{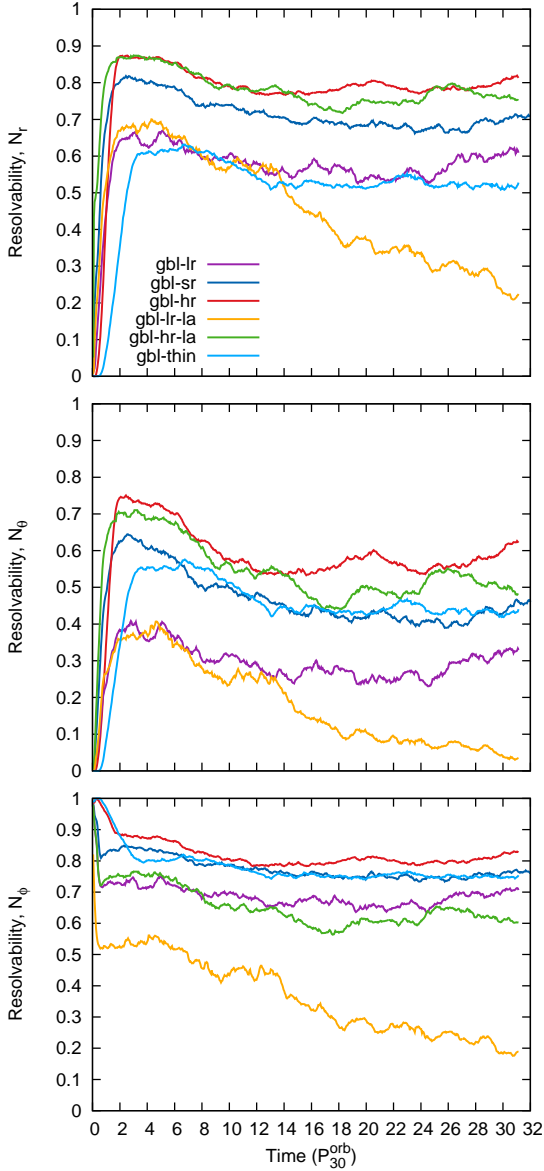}} \\
   \end{tabular}
   \caption{Resolvability (see Eq~\ref{eqn:resolvability}) of the MRI
     in the global simulations in the $r$ (upper), $\theta$ (middle),
     and $\phi$-directions (lower).}
    \label{fig:resolvability}
  \end{center}
\end{figure}

\begin{figure}
  \begin{center}
    \begin{tabular}{c}
\resizebox{75mm}{!}{\includegraphics{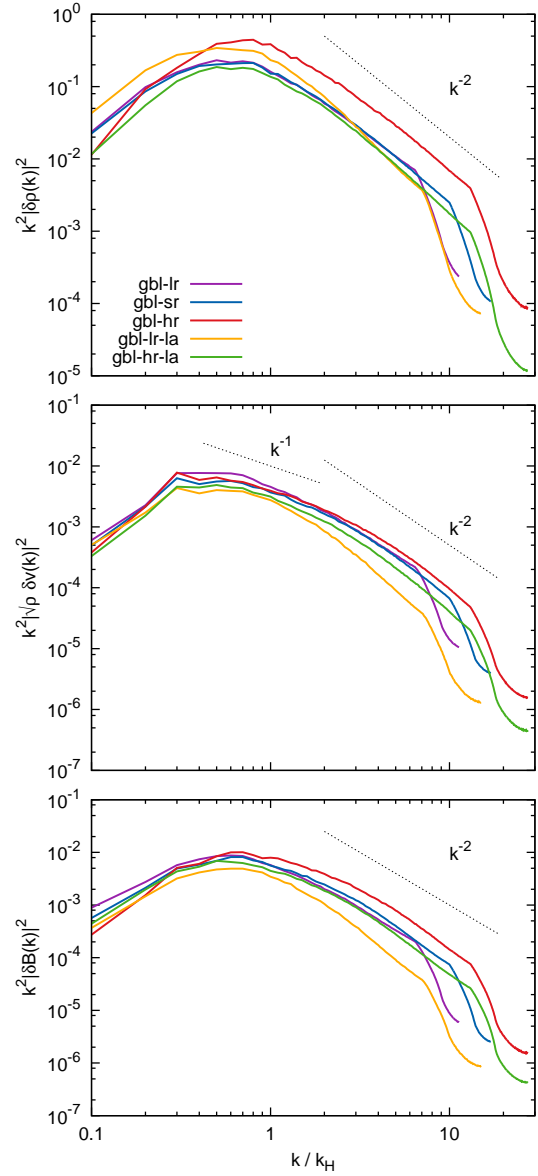}} \\
   \end{tabular}
   \caption{Angle-averaged energy spectrum calculated from
     time-averaged simulation data showing density (top), kinetic
     energy (middle), and magnetic energy (lower). The dotted lines
     are representative power-law slopes (see
     \S~\ref{sec:saturation}). The horizontal axis is in units of
     $k_{\rm H}=2\pi/\langle H \rangle$.}
    \label{fig:power_spectra}
  \end{center}
\end{figure}

\begin{figure}
  \begin{center}
    \begin{tabular}{c}
\resizebox{75mm}{!}{\includegraphics{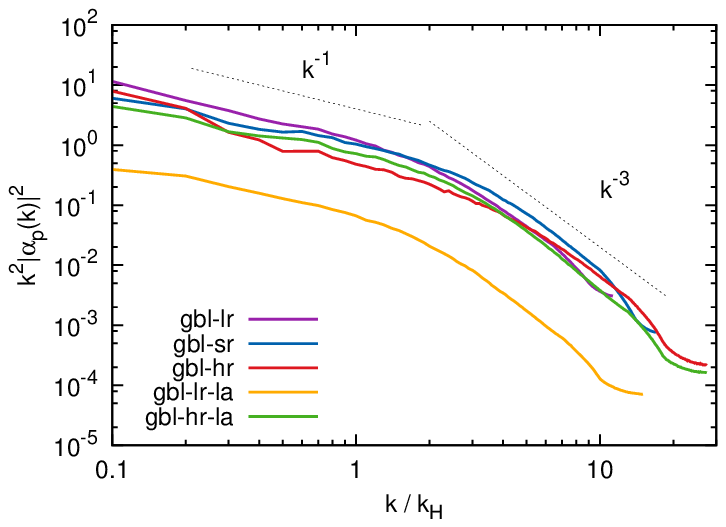}} \\
   \end{tabular}
   \caption{Angle-averaged energy spectrum for the
     stress-to-gas-pressure ratio, $|\alpha_{\rm p}({\bf k})|^2$,
     calculated from time-averaged simulation data. The horizontal
     axis is in units of $k_{\rm H}=2\pi/\langle H \rangle$.}
    \label{fig:alphap_pds}
  \end{center}
\end{figure}

\section{The quasi-steady state}
\label{sec:saturation}

Following the initial transient phase of evolution, the disk settles
into a QSS. In this section we examine the characteristics of this
state for the different simulations. We list time and volume averaged
parameter values pertaining to the steady-state turbulence in
Table~\ref{tab:global_models}.

\begin{figure}
  \begin{center}
    \begin{tabular}{c}
\resizebox{80mm}{!}{\includegraphics{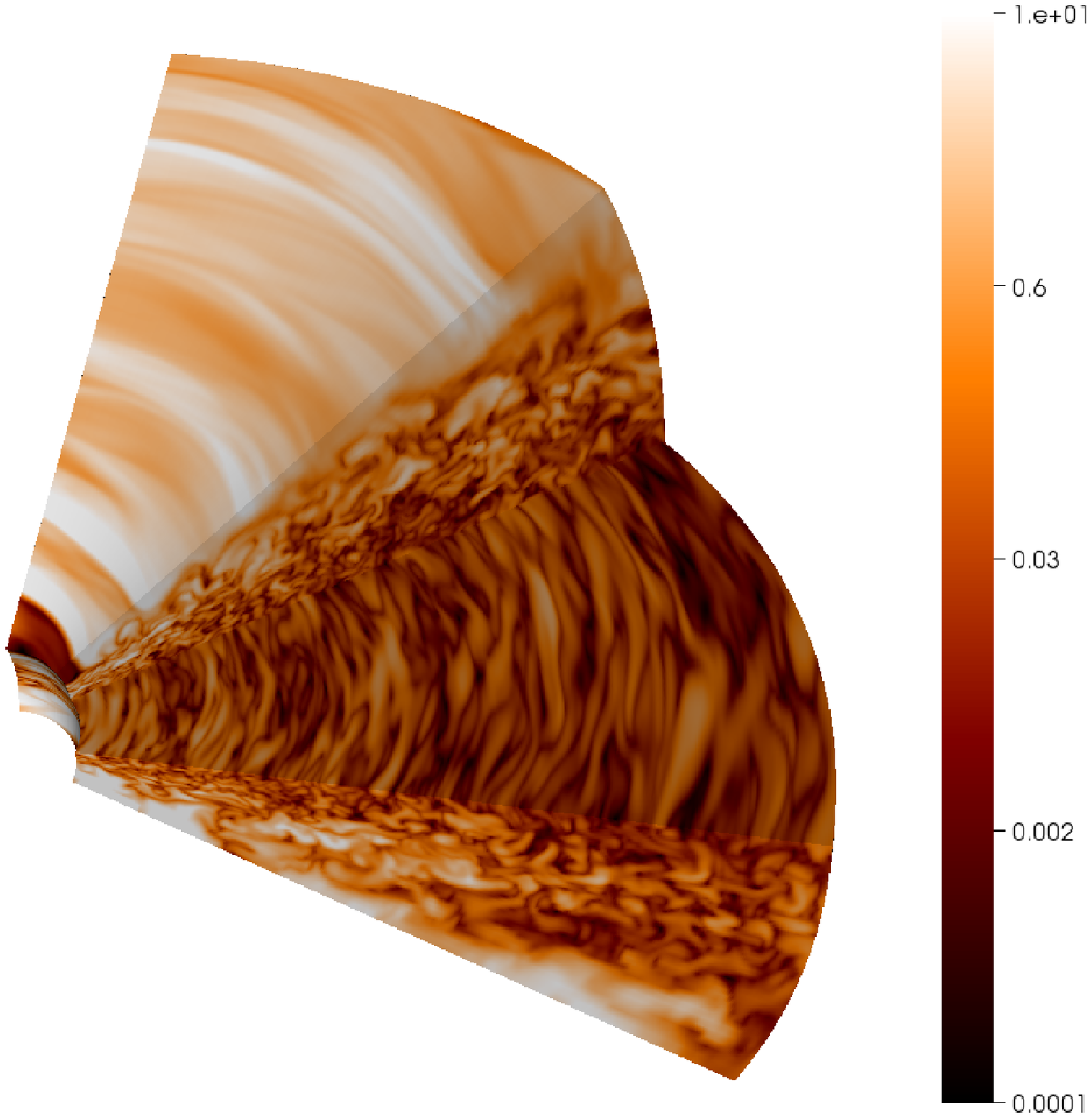}} \\
\resizebox{80mm}{!}{\includegraphics{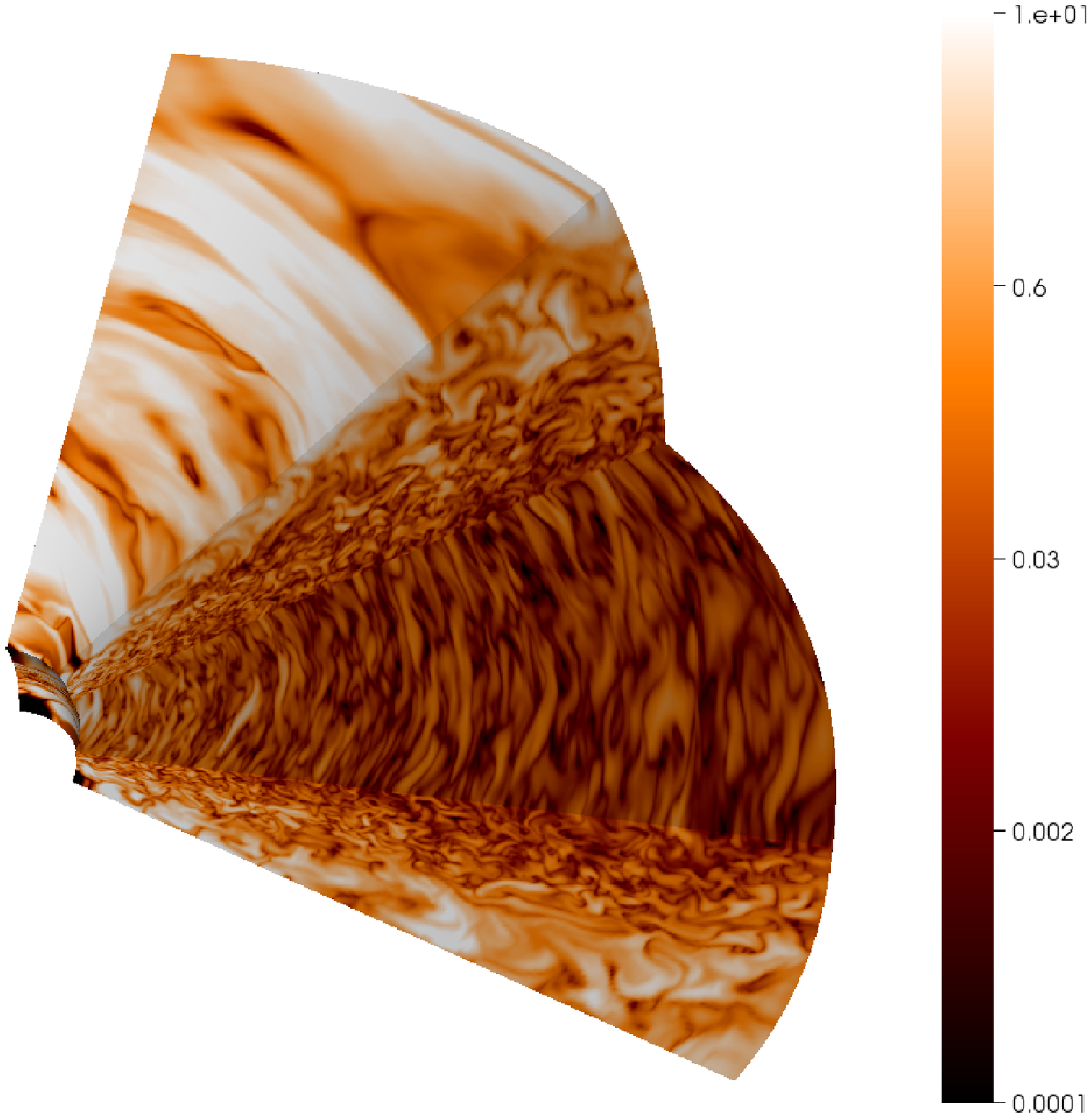}} \\
\resizebox{80mm}{!}{\includegraphics{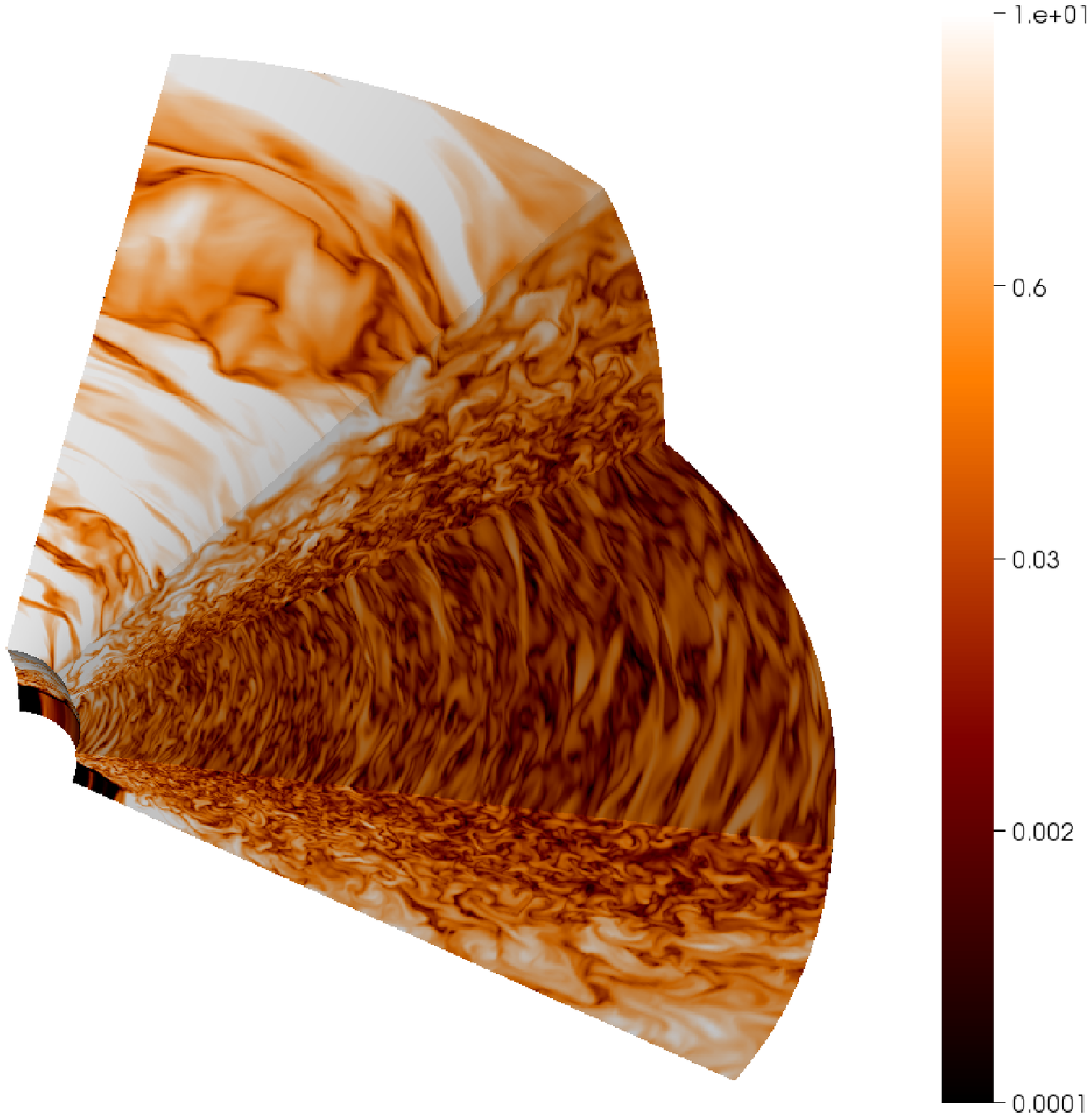}} \\
 \end{tabular}
  \caption{3D volume rendering showing the ratio of magnetic pressure
    to gas pressure ($\beta^{-1}$) for models gbl-lr (top), gbl-sr
    (middle), and gbl-hr (lower). A wedge has been excised from the
    upper hemisphere of the disk to expose the disk mid-plane.}
    \label{fig:snapshots}
  \end{center}
\end{figure}

\subsection{Resolution dependence}

\subsubsection{Convergence: gbl-lr, gbl-sr, and gbl-hr}

The volume averaged stress normalised to gas pressure,
\linebreak$\langle\alpha_{\rm P}\rangle$, displays a dependence on
resolution in the magnitude of the transient peak at $t\sim 2~P^{\rm
  orb}_{30}$ (Fig.~\ref{fig:alpha}). Following this,
$\langle\alpha_{\rm P}\rangle$ steadily declines until $t\sim
12~P^{\rm orb}_{30}$, at which point the curves level-off and a
quasi-steady state is reached. We find a time-averaged value during
the quasi-steady state of $\overline{\langle\alpha_{\rm P}\rangle}
\simeq 0.04$ for models gbl-lr, gbl-sr, and gbl-hr, indicating
convergence with resolution. At the point of convergence,
$\overline{\langle\alpha_ {\rm
    P}\rangle}\thinspace\overline{\langle\beta_{\rm
    d}\rangle}\simeq0.6$, where $\beta_{\rm d}$ is the disk body
plasma-$\beta$. This is in agreement with, but slightly higher than,
the relation found for unstratified shearing-box simulations
\citep{Sano:2004, Blackman:2008, Guan:2009}. The volume averaged
Maxwell stress normalised to the magnetic pressure,
$\langle\alpha_{\rm M}\rangle$ converges at a value of 0.42 (lower
panel of Fig.~\ref{fig:alpha}), consistent with previous high
resolution stratified shearing box \citep{Hawley:2011, Simon:2012} and
global disk models \citep{Parkin:2013}.

Convergence in $\overline{\langle\alpha_{\rm P}\rangle}$ is coincident
with convergence in the resolvability (Eq~\ref{eqn:resolvability}) -
the ability of the numerical grid to resolve the fastest growing MRI
modes.  Examining Fig.~\ref{fig:resolvability}, one sees that the
$\phi$-direction is the best resolved, followed by the radial
direction, and then the $\theta$-direction. At the resolution of model
gbl-sr, $N_{\rm r}$ is clearly higher than $N_{\theta}$, suggesting
that convergence in global models is tied to the radial magnetic
field. That convergence with resolution is more readily achievable for
the radial magnetic field is illustrated by the relative magnetic
field strengths: $\overline{\langle\beta_{\rm r}\rangle} \simeq 128$,
$\overline{\langle\beta_{\theta}\rangle} \simeq 361$, and
$\overline{\langle\beta_{\phi}\rangle} \simeq 17$. The converged value
for $N_{\phi}$ is roughly 0.8, consistent with some fraction of the
disk having weak magnetic fields (due to zero-net flux dynamo
oscillations) which have corresponding $\lambda_{\rm MRI}$ values
which are below the simulation resolution. Poloidal magnetic fields in
our simulations appear relatively strong, with our models returning
$\overline{<B_{\rm r}^2>/<B_{\phi}^2>}=0.13$ and
$\overline{<B_{\theta}^2>/<B_{\phi}^2>}=0.05$ compared to respective
values of $\sim 0.08$ and $\sim 0.02$ for the highest resolution model
in \cite{Hawley:2011}. We attribute this difference to the higher
resolution used in our models.

Power spectra computed for density, kinetic energy, and total magnetic
energy perturbations are shown in Fig.~\ref{fig:power_spectra}. A
perturbation for a variable q is calculated by subtracting the
azimuthal average, such that $\delta q = q - [q]$. Perturbations are
used to reduce the influence of large scale radial and vertical
gradients on the resulting power spectra. The position of the low
wavenumber turnover between models gbl-lr, gbl-sr, and gbl-hr is
consistent at $k/k_{\rm H}\simeq 0.5$, illustrating that the
resolution of the largest physical structures is converged. The slope
of the magnetic energy power spectrum is approximately $k^{-2}$. This
apparent constant power-law slope suggests a self-similar transfer of
energy from large to small scales which may indicate an inertial
cascade, although it may also be due to the injection of energy by the
MRI at all realisable scales \citep{Fromang:2007a}. The power spectra
for magnetic energy and kinetic energy exhibit very similar shapes. On
further inspection one sees that magnetic energy is slightly larger
than kinetic energy on length scales of roughly a disk scale-height
($0.4\ltsimm k/k_{\rm H} \ltsimm 5$, where $k_{\rm H}=2\pi/\langle H
\rangle$), whereas they are approximately equal on the smallest length
scales ($k/k_{\rm H}\gtsimm 5$). This differs from the stratified
shearing-box simulations presented by \cite{Johansen:2009}, for which
kinetic energy was found to dominate over magnetic energy at all but
the very largest scales in the box. Comparing to the global models of
\cite{Beckwith:2011}, we note that the low wavenumber turnover for
magnetic and kinetic energy arises at a similar value (they find
$k/k_{\rm H}\sim0.3$). However, there is a considerable difference in
the amplitudes of kinetic and magnetic energy fluctuations, where
\cite{Beckwith:2011} find the latter to be an order of magnitude lower
than the former. The source of this difference is unclear. However,
there are number of differences between our approach and that used by
\cite{Beckwith:2011} in the calculation of the Fourier transforms and
the related power spectra. In calculating the value of a fluctuating
quantities $\delta Q$ we have adopted a straightforward approach of
subtracting an azimuthally averaged value of $Q$, whereas
\cite{Beckwith:2011} fit a two-dimensional distribution in the radial
and vertical directions, which they then subtract to determine $\delta
Q$. Another major difference is that we define a conventional
spherical Fourier transform through Eq~(\ref{eqn:ft}), whereas
\cite{Beckwith:2011} construct azimuthal averages of fluctuating
quantities, define a normalized measure of spatial fluctuations in the
$(r,\theta)$ coordinates and then define a Fourier transform in $r$
and $\theta$ treating $r$ and $\theta$ as pseudo-Cartesian coordinates 
(their equation~(15)). A comparison of these two approaches and the
implications for comparing computed accretion disk spectra with one
another and also with textbook spectra for homogeneous turbulence, is
beyond the scope of this paper.

The power spectra all display a pronounced turn-over at high
wavenumber, which depends on resolution, and which we interpret as the
dissipation scale. Indeed the morphology of the steep, but slightly
curved, step at the high wavenumber end of the magnetic energy power
spectrum is indicative of a resolved separation between the Ohmic and
viscous dissipation scales \citep[see, e.g.,][]{Kraichnan:1967}.

Irrespective of resolution, most of the power in $\alpha_{\rm P}$ is
on the largest length scales (i.e. at low wavenumber -
Fig.~\ref{fig:alphap_pds}). In fact, the relatively flat slope to the
$\alpha_{\rm P}$ power spectrum indicates that a large amount of power
is also contained in moderate length scales. The slope of the power
spectrum changes at $k/k_{\rm H} \sim3$, becoming steeper, and
indicating that smaller length scales contribute considerably less to
the global stress. Therefore, although magnetic field correlation
lengths demonstrate that MRI-driven turbulence is localised
\citep{Guan:2009}, we find evidence for angular momentum transport
being dominated by larger length scales, of size $\gtsimm \langle H
\rangle$.

Increasing the simulation resolution permits structure to occupy
smaller spatial scales. This is illustrated by the simulation
snapshots of $\beta^{-1}$ shown in Fig.~\ref{fig:snapshots}. As one
progresses to higher resolution through models gbl-lr, gbl-sr, and
gbl-hr the size of structures get progressively finer. Also, contrasts
in the magnetic energy, which are particularly noticeable in the
coronal region, become sharper at higher resolution. This equates to
an increase in $\nabla \times B$ with resolution, which we examine in
more detail in \S~\ref{sec:control_volume}.

In summary, convergence is achieved for a resolution of 12-51
cells/$H$ in radius, 27 cells/$H$ in the $\theta$-direction, and 12.5
cells/$H$ in the $\phi$-direction (model gbl-sr). This is considerably
below the 64-128 cells/$H$ required for convergence in stratified
shearing box simulations found by \cite{Davis:2010}, whereas the
vertical resolution is comparable to the 25 cells/$H$ necessary to
produce sustained turbulence in the models of \cite{Fromang:2006} and
\cite{Flock:2011}. In \S~\ref{sec:convergence} we provide an
explanation for this dramatic difference.

\subsubsection{Influence of $\phi-resolution$: gbl-lr-la and gbl-hr-la}

When the azimuthal field is under-resolved, turbulent activity dies
out, as discussed by \cite{Fromang:2006}, \cite{Flock:2011}, and
\cite{Parkin:2013}. This effect can be seen in the stresses and
resolvabilities computed for model gbl-lr-la (a lower azimuthal
resolution variant of gbl-lr - see Table~\ref{tab:model_list}), which
we plot in Figs.~\ref{fig:alpha} and
\ref{fig:resolvability}. Repeating this experiment at higher
resolution (models gbl-hr and gbl-hr-la), one finds that even though
the azimuthal field is barely-resolved (8 cells/$H$ in the azimuthal
direction for gbl-hr-la), only a slightly lower
$\overline{\langle\alpha_{\rm P}\rangle}$ value is
obtained. Therefore, we find a similar dependence on azimuthal
resolution to that discussed by \cite{Hawley:2011}, although this
dependence appears to become less pronounced at higher resolution, and
this is possibly due to compensation by the poloidal grid
resolution. This indicates that low azimuthal resolution can, to some
extent, be compensated for by higher poloidal resolution. However,
based on these results it would seem advisable to adopt an aspect
ratio close to unity. We also note that our $\alpha_{\rm P}$ and
$\alpha_{\rm M}$ values are higher than in the models of
\cite{Beckwith:2011} and \cite{Hawley:2011}. We attribute this to the
higher simulation resolution and lower cell aspect ratio used in our
models - \citep[see also the discussion in][]{Fromang:2006,
  Flock:2011, Parkin:2013}.

As discussed in the previous section, the power spectra in
Fig.~\ref{fig:power_spectra} display a turn-over at high wavenumber
corresponding to the dissipation scale. All simulations presented in
this paper rely on numerical dissipation, hence one may anticipate
that numerical resolution sets this scale and adopting a lower
resolution in a certain direction may shift the dissipation scale to
lower wavenumbers. Comparing the curves for models gbl-lr and
gbl-lr-la in Fig.~\ref{fig:power_spectra}, the slope in the magnetic
energy power spectrum is steeper for gbl-lr-la in the wavenumber range
$1\ltsimm k/k_{\rm H} \ltsimm 6$. This steeper slope is readily
understood as a consequence of under-resolving the fastest growing MRI
modes in the $\phi$-direction - energy cannot be injected by the MRI
if the mode-growth is not resolved. In relation to the discussion in
the previous section, this implies that the power-law slope in the
magnetic energy power spectrum of model gbl-lr, for example, is a
consequence of magnetic energy injection by the MRI and not solely due
to an inertial cascade of energy from large to small scales,
consistent with results presented by \cite{Fromang:2007a} and
\cite{Johansen:2009} which illustrate the driving of magnetic energy
to smaller scales by the MRI.

\subsection{Other factors which might affect the saturated state}

\subsubsection{Gas pressure dependence: gbl-thin}

Based on a suite of unstratified shearing-box simulations,
\cite{Sano:2004} have reported a dependence of the QSS stress level on
the gas pressure. To examine whether this dependence exists in global
simulations, one can utilise simulations with different aspect ratios
because $H/R \propto c_{\rm s} \propto \sqrt{P}$. Models gbl-sr and
gbl-thin\footnote{Model gbl-thin was previously presented in
  \cite{Parkin:2013} as gbl-$m10+$ and further analysis can be found
  there-in.}, which feature $H/R=0.1$ and 0.05, respectively, and an
identical number of cells per scale-height in the vertical and
azimuthal directions (Table~\ref{tab:model_list}), return comparable
values of $\overline{\langle\alpha_{\rm P}\rangle}$, with a slightly
higher value for the latter (Table~\ref{tab:global_models}). Based on
the two disk aspect ratios we have explored, there is a weak
dependence of $\langle\alpha_{\rm P}\rangle$ on gas pressure. This
lack of dependence stems from the similarity in values for the disk
body plasma-$\beta$, $\beta_{\rm d}$
(Table~\ref{tab:global_models}). Essentially, even though gas pressure
is higher in gbl-sr compared to gbl-thin, the relative strength of the
Maxwell stresses is very similar.

\subsubsection{Initial magnetic field strength}

The independence of the saturated state on the initial field strength
has been demonstrated by \cite{Sano:2004} (zero-net flux, unstratified
shearing-boxes), \cite{Guan:2011} (stratified shearing-boxes), and
\cite{Hawley:2011} (global stratified disks). In all cases, the
dissipation, or expulsion, of the initial magnetic field configuration
disconnects its influence from the QSS. 

\subsubsection{Initial perturbation to the disk}

The growth rate of the non-axisymmetric MRI depends on the wavenumber
of the initial perturbation \citep{BH92}. Faster magnetic field growth
arises for higher wavenumbers provided they are
sub-critical. Therefore, the growth rate of the stresses in the disk
will be higher for higher wavenumber perturbations. \cite{Parkin:2013}
examined this in the context of global disks, finding that
irrespective of the initially excited MRI mode, the onset of
non-linear (turbulent) motions in the disk erases the initial
perturbation and leads to a statistically similar saturated state.

\subsection{Astrophysical implications}

Simulations gbl-lr, gbl-sr, and gbl-hr converge at
$\overline{\langle\alpha_{\rm P}\rangle}\simeq 0.04$ which should be
compared with values of $\sim0.1-0.3$ commonly derived from relaxation
times in post-outburst cataclysmic variables (CVs)\citep{Smak:1999,
  King:2007, Kotko:2012}. Although a discrepancy exists, we note that
our converged values are consistent with those for isolated AGN disks
\citep{Starling:2004}. Attaining higher values for
$\overline{\langle\alpha_{\rm P}\rangle}$ may, therefore, require the
influence of a companion star to be included in
simulations. Alternatively, large-scale (net vertical flux) magnetic
fields and/or large magnetic Prandtl numbers have been shown to yield
larger stresses \citep{Lesur:2007,Fromang:2013, Bai:2013, Lesur:2013}.

\section{Control volume analysis}
\label{sec:control_volume}

In order to understand the global characteristics of our model
accretion disks, we have preformed a control volume analysis of the
magnetic energy budget. This involves evaluating terms in the magnetic
energy equation integrated over a specific volume, and for this
purpose we choose the disk body (defined in
\S~\ref{subsec:diagnostics}). The boundaries of this control volume
are open in the radial and vertical directions, and periodic in the
azimuthal direction.

Previous shearing box simulations show a common characteristic of a
consistent power on large scales when convergence is achieved
\citep{Simon:2009, Davis:2010}. Thus the large scale power in the
turbulent spectrum is important when assessing the convergence
properties of accretion disk simulations. In turbulent gases (and in
our simulations) most of the energy is contained in the largest scales
so that our volume-integrated approach to the energy budget is useful
in understanding the characteristics of the low wave number part of
the spectrum.

Another important aspect of the control volume analysis is that it
provides a guide for constraining the various terms describing
production, advection, volumetric changes and dissipation in analytic
modeles of accretion disks \citep[e.g.][]{Balbus:1999,
  Kuncic:2004}. For example, this analysis informs us whether the
vertical advection of turbulent energy is important compared to the
production of turbulence in magnetised disks.

\begin{figure}
  \begin{center}
    \begin{tabular}{c}
\resizebox{80mm}{!}{\includegraphics{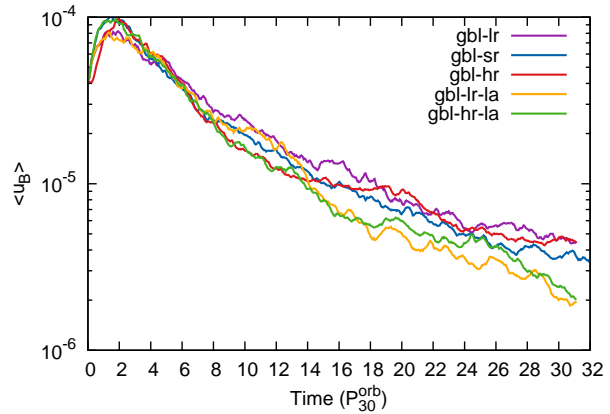}} \\
   \end{tabular}
   \caption{Volume averaged magnetic energy, $\langle u_{\rm
       B}\rangle$, as a function of time.}
    \label{fig:uB}
  \end{center}
\end{figure}

\begin{figure}
  \begin{center}
    \begin{tabular}{c}
\resizebox{80mm}{!}{\includegraphics{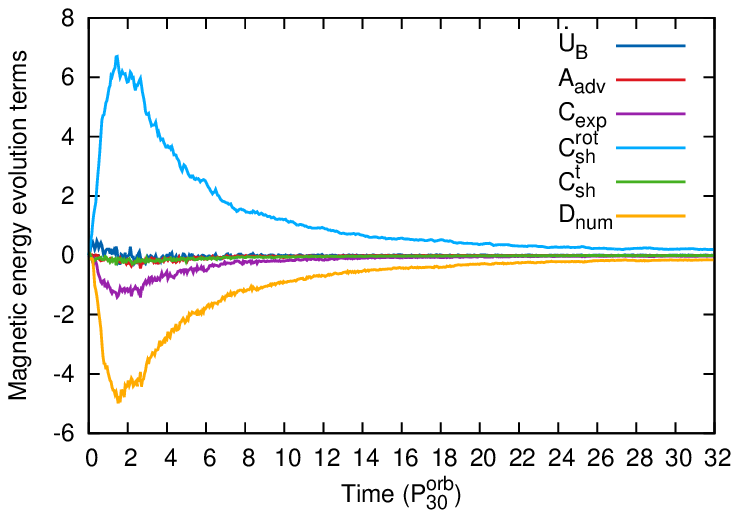}} \\
\resizebox{80mm}{!}{\includegraphics{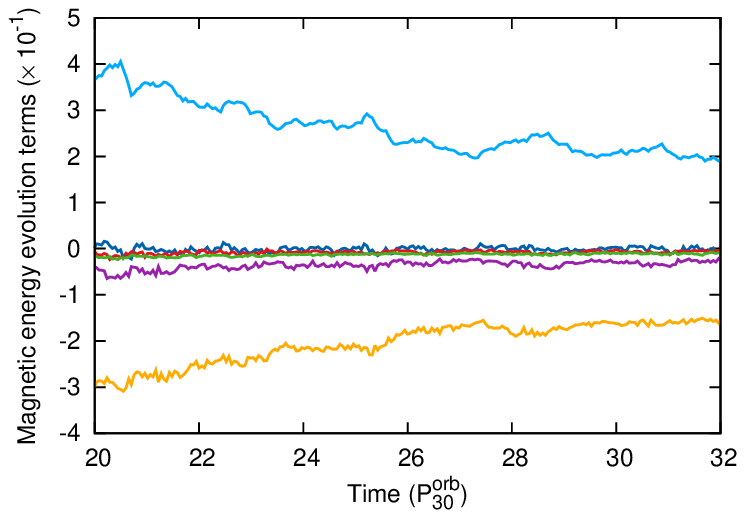}} \\
  \end{tabular}
  \caption{Comparison of terms pertaining to the control volume
    analysis for model gbl-sr (see \S~\ref{subsec:mag_energy}). The
    upper and lower panels show results for time intervals 0-32 and
    $20-32~P_{30}^{\rm orb}$, respectively. Rates of change of energy
    are plotted in units of $1/P_{30}^{\rm orb}$ - to convert to code
    units divide the values by $P_{30}^{\rm orb}=964$.  Note the
    difference in scale between the plots.}
    \label{fig:control_vol}
  \end{center}
\end{figure}

\subsection{Magnetic energy evolution}
\label{subsec:mag_energy}

To set the scene, we first examine the control volume averaged
magnetic energy, $\langle u_{\rm B}\rangle$ (Fig.~\ref{fig:uB}). The
curves follow the general morphology of a rapid rise in $u_{\rm B}$
during the initial transient phase, followed by a similarly rapid fall
in magnetic energy which gradually flattens out as the quasi-steady
state ($\overline{\partial \langle\alpha_{\rm P}\rangle/\partial t}
\rightarrow 0$) is reached. For gbl-sr, for example, the quasi-steady
state is reached after $t\simeq 14~P^{\rm orb}_{30}$. Subsequently,
there is a slow, but steady, decrease in magnetic energy.

We begin with the magnetic field induction equation, to which we add a
term for {\it numerical} resistivity, $\eta_{\rm num}$, such that
Eq~(\ref{eqn:induction}) now reads,
\begin{equation}
\frac{\partial\bf B}{\partial t} = \nabla \times ({\bf v \times
  B}) + \eta_{\rm num}\nabla^2 {\bf B}. \label{eqn:induction2}
\end{equation}
The motivation for introducing $\eta_{\rm num}$ will become clear in
the remainder of the paper. For now we merely note that the truncated
order of accuracy of numerical finite volume codes (such as the
{\sevensize PLUTO} code used in this investigation) brings with it a
truncation error which we interpret as a numerical resistivity and
which we model with the additional Ohmic term in
Eq~(\ref{eqn:induction2}). Taking the scalar product of ${\bf B}$ with
Eq~(\ref{eqn:induction2}) and re-arranging terms gives,
\begin{equation}
\frac{\partial u_{\rm B}}{\partial t} = B_{\rm i}B_{\rm j}s_{\rm ij} -
\frac{1}{3}u_{\rm B}v_{\rm k,k} - \frac{\partial}{\partial x_{\rm
    j}}(u_{\rm B} v_{\rm j}) + \eta_{\rm num}B_{\rm i}\nabla^2 B_{\rm
  i}, \label{eqn:mag_energy1}
\end{equation}
where $u_{\rm B}=|B|^2/2$, the fluid shear tensor,
\begin{equation}
  s_{\rm ij} = \frac{1}{2}(v_{\rm i,j} + v_{\rm j,i} - \frac{2}{3}
  \delta_{\rm ij} v_{\rm k,k}), \label{eqn:fluidstress}
\end{equation}
and the Maxwell stress tensor is given by Eq~(\ref{eqn:maxstress}),
and a subscript comma denotes partial differentiation. Next we expand
${\bf v} = {\bf v}^{\rm t} + {\bf v}^{\rm rot}$, where $v^{\rm t}$ is
the perturbed velocity field in the rotating frame and,
\begin{equation}
  {\bf v}^{\rm rot} = v^{\rm rot}\hat{\phi} = [v_{\phi}]\hat{\phi}, \label{eqn:vsh}
\end{equation}
is the azimuthally averaged rotational velocity. This step allows the
respective contributions to the terms in Eq~(\ref{eqn:mag_energy1})
from the mean background disk rotation and the perturbed velocity
field (in the rotating frame) to be inspected. Substituting
Eq~(\ref{eqn:vsh}) into Eq~(\ref{eqn:mag_energy1}) and integrating
over a control volume $V$ with bounding surface $S$, and using the
relation,
\begin{equation}
  B_{\rm i}B_{\rm j}s_{\rm ij} - \frac{1}{3}u_{\rm B}v_{\rm k,k} =
  B_{\rm i}B_{\rm j}v_{\rm i;j} - u_{\rm B}v_{\rm k,k},
\end{equation}
to separate shear and expansion terms (where a subscript semi-colon
indicates a covariant derivative) one arrives at,
\begin{eqnarray}
  \dot{U}_{\rm B} = C^{\rm rot}_{\rm sh} + C^{\rm t}_{\rm sh} + C_{\rm exp} + A_{\rm adv} + D_{\rm num}, \label{eqn:mag_energy2}
\end{eqnarray}
where,
\begin{eqnarray}
  \dot{U}_{\rm B} &=& \int \frac{\partial u_{\rm B}}{\partial t}dV, \label{eqn:UB}\\
  C^{\rm rot}_{\rm sh} &=&\int B_{\rm i}B_{\rm j}v^{\rm rot}_{\rm i;j} dV, \\
  C^{\rm t}_{\rm sh} &=&\int B_{\rm i}B_{\rm j}v^{\rm t}_{\rm i;j} dV, \\
  C_{\rm exp} &=&- \int u_{\rm B}v^{\rm t}_{\rm k,k} dV, \\
  A_{\rm adv} &=&- \oint u_{\rm B}v^{\rm t}_{\rm j}n_{\rm j}dS, \label{eqn:Aadv}\\
  D_{\rm num} &=&-\eta_{\rm num}\left[\oint \frac{\partial M^{\rm
        B}_{\rm ij}}{\partial x_{\rm i}}n_{\rm j}dS + \int |j|^2 dV\right], \label{eqn:Dnum}
\end{eqnarray}
and where the current density, $j_{\rm i}=[\nabla \times B]_{\rm i}$,
and the Maxwell stress tensor, $M^{\rm B}_{\rm ij}$, is given by
(\ref{eqn:maxstress}). All terms featuring in
Eqs~(\ref{eqn:mag_energy2})-(\ref{eqn:Dnum}) are exact and can be
explicitly calculated from the simulation data. The numerical
resistivity, $\eta_{\rm num}$, is estimated by solving
Eq~(\ref{eqn:mag_energy2}) for $D_{\rm num}$ and then solving
Eq~(\ref{eqn:Dnum}) for $\eta_{\rm num}$. To maintain consistency with
the third-order spatial reconstruction used in the simulations, we
compute terms appearing in
Eqs~(\ref{eqn:mag_energy2})-(\ref{eqn:Dnum}) to third-order accuracy
using reconstruction via the primitive function \citep[see][for
further details]{Colella:1984, Laney:1998}.

Before proceeding to the results of the control volume analysis, a
brief description of the terms and their respective meaning is
worthwhile. The volume integrated rate of change of magnetic energy is
given by $\dot{U}_{\rm B}$. $C^{\rm rot}_{\rm sh}$ and $C^{\rm t}_{\rm
  sh}$ are the production of magnetic energy by the shear in the mean
disk rotation and the turbulent velocity field, respectively. $C_{\rm
  exp}$ corresponds to changes in magnetic energy due to expansion in
the gas. $A_{\rm adv}$ is a surface term for the advection of magnetic
energy in/out of the control volume by the turbulent velocity
field. There are contributions to the surface integrals from the
radial and $\theta$-direction - periodic boundaries in the azimuthal
direction lead to a cancellation, and thus no contribution from those
surfaces. (Note that the term $\oint u_{\rm B}v^{\rm rot}_{\rm
  j}n_{\rm j} dS$ vanishes because of the periodic boundary conditions
in the azimuthal direction, therefore the mean disk rotation does not
advect magnetic energy in/out of the control volume.) Finally, $D_{\rm
  num}$ corresponds to numerical dissipation. We note that the value
of $D_{\rm num}$ is exact, as it is merely the remainder required to
balance the magnetic energy equation
(Eq~\ref{eqn:mag_energy2}). However, our determination of $\eta_{\rm
  num}$ from $D_{\rm num}$ is not exact in view of our {\it assumed}
Ohmic form for the numerical resistive term in
Eq~(\ref{eqn:induction2}). Nevertheless, we consider this estimate to
be indicative of the actual numerical resistivity.

In Fig.~\ref{fig:control_vol} we plot the results of applying the
control volume analysis to model gbl-sr. One immediately notices that
magnetic energy production is dominated by $C^{\rm rot}_{\rm sh}$
\citep[see also discussion in][]{Kuncic:2004} and removal is
predominantly via numerical dissipation, $D_{\rm num}$. There is
non-negligible magnetic energy removal by divergence in the velocity
field ($C_{\rm exp}$). Examining the directional contributions to this
term, one finds roughly equal magnitudes for the $r$, $\theta$, and
$\phi$ components. However, the poloidal contributions ($r,\theta$)
are expansions, which remove magnetic energy, whereas the azimuthal
contribution is compressive, thus being a source of magnetic
energy. Flow divergence impacting on magnetic field evolution has also
been observed in stratified shearing-box simulations by
\cite{Johansen:2009}. The turbulent velocity field does not contribute
greatly to magnetic energy production, as is demonstrated by the
comparably small values for the $C^{\rm t}_{\rm sh}$ curve. A
negligible amount of magnetic energy appears to be advected out of the
volume in the radial and vertical directions as shown by the curve for
$A_{\rm adv}$, consistent with stable magnetically buoyant (Parker)
modes within $|z|<2H$ \citep{Shi:2010}. Therefore, although
``butterfly'' diagrams indicate quasi-periodic vertical magnetic field
expulsion \citep[e.g.][]{Gressel:2010}, it would seem that a much
greater amount of energy is dissipated within the disk body. The rate
of change of magnetic energy is relatively small compared to magnetic
energy production by $C^{\rm rot}_{\rm sh}$ and dissipation by $D_{\rm
  num}$. Computing a time-averaged value between orbits 12-32, we find
$\overline{\dot{U}_{\rm B}}=-3.5\times10^{-6}$. Therefore, although
$\langle\alpha_{\rm P}\rangle$ exhibits quasi-steady behaviour,
$\langle u_{\rm B}\rangle$ is continually declining, but at a
constantly decreasing rate.

\begin{figure}
  \begin{center}
    \begin{tabular}{c}
\resizebox{80mm}{!}{\includegraphics{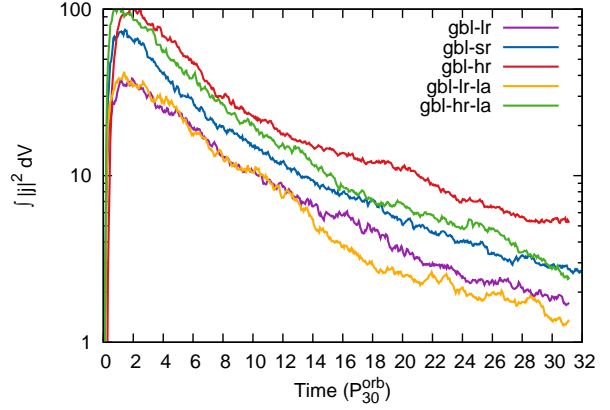}} \\
   \end{tabular}
   \caption{Integral of the current density squared, $|j|^{2}$, over
     the control volume. This term indicates the level of turbulent
     activity in the magnetic field and is the dominant contributor to
     $D_{\rm num}$.}
    \label{fig:curlB2}
  \end{center}
\end{figure}

Examining the dissipation term, $D_{\rm num}$, in more detail, one
finds that the first term in square brackets on the RHS of
Eq~(\ref{eqn:Dnum}) is considerably smaller than the second. This
shows that dissipation is primarily powered by the current
density\footnote{The link between the turbulent magnetic field and the
  current density bears strong similarities to the that between the
  velocity field and the vorticity, $\omega_{\rm i}=[\nabla \times
  v]_{\rm i}$.}, $|j|$. In Fig.~\ref{fig:curlB2} we show $\int |j|^2
dV$. There is a striking similarity between the morphology of the
curves in this plot with those for $\langle u_{\rm B}\rangle$
(Fig.~\ref{fig:uB}), suggesting an intimate link between the evolution
of the magnetic energy, dissipation driven by a turbulent magnetic
field, and magnetic energy production (also demonstrated by $C^{\rm
  rot}_{\rm sh}$ and $D_{\rm num}$ in
Fig.~\ref{fig:control_vol}). Comparing time-averaged values for
$D_{\rm num}$ from models gbl-lr, gbl-sr, and gbl-hr, we find very
little difference. Therefore, as convergence with resolution is
achieved for $\overline{\langle\alpha_{\rm P}\rangle}$, the level of
dissipation also converges. We elaborate on the above points in
\S~\ref{sec:convergence} in the context of a unified description for
the observed evolution in accretion disk simulations.

The formulation of the magnetic energy equation used in
Eq~(\ref{eqn:mag_energy2}) allows one to distinguish the contributions
from shearing and expansions in the disk. However, with a view to
understanding the influence of the boundary conditions for the control
volume on the magnetic energy, and on the power injected by Maxwell
stresses and Lorentz forces, an alternative formulation may be
used. To this end we re-cast Eq~(\ref{eqn:mag_energy2}) as:
\begin{equation}
  \dot{U}_{\rm B} = C^{\rm rot}_{\rm Lor} + C^{\rm t}_{\rm Lor} +
  A^{\rm rot}_{\rm MS} + A^{\rm t}_{\rm MS}+ A_{\rm adv} + D_{\rm num}, \label{eqn:mag_energy3}
\end{equation}
where, 
\begin{eqnarray}
  C^{\rm rot}_{\rm Lor} &=& - \int v^{\rm rot}_{\rm i} \frac{\partial M^{\rm B}_{\rm
      ij}}{\partial x_{\rm j}} dV = - \int v^{\rm rot}_{\rm i} F^{\rm L}_{\rm i} dV,\\
  C^{\rm t}_{\rm Lor} &=& - \int v^{\rm t}_{\rm i} \frac{\partial M^{\rm B}_{\rm
      ij}}{\partial x_{\rm j}} dV = - \int v^{\rm t}_{\rm i} F^{\rm
    L}_{\rm i} dV,\\
  C_{\rm Lor} &=& C^{\rm rot}_{\rm Lor} + C^{\rm t}_{\rm Lor},\label{eqn:CLor}\\
  A^{\rm rot}_{\rm MS} &=& \oint M^{\rm B}_{\rm ij} v^{\rm rot}_{\rm j} n_{\rm i}
  dS, \\
  A^{\rm t}_{\rm MS} &=& \oint M^{\rm B}_{\rm ij} v^{\rm t}_{\rm j}
  n_{\rm i} dS,\\
A_{\rm MS} &=& A^{\rm rot}_{\rm MS} + A^{\rm t}_{\rm MS},\label{eqn:AMax} 
\end{eqnarray}
and where $\dot{U}_{\rm B}$, $A_{\rm adv}$, and $D_{\rm num}$ are
given by Eqs~(\ref{eqn:UB}), (\ref{eqn:Aadv}), and (\ref{eqn:Dnum}),
respectively. The rates of work done within the control volume by the
Lorentz force, $F^{\rm L}$, in combination with the mean disk
rotation, $v^{\rm rot}$, and the turbulent velocity field (in the
rotating frame), $v^{\rm t}$, are given by $C^{\rm rot}_{\rm Lor}$ and
$C^{\rm t}_{\rm Lor}$, respectively. Similarly, the rates of work done
on the surfaces of the control volume by combinations of the Maxwell
stresses and $v^{\rm rot}$ and $v^{\rm t}$ are, respectively, given by
$A^{\rm rot}_{\rm MS}$ and $A^{\rm t}_{\rm MS}$.

\begin{figure}
  \begin{center}
    \begin{tabular}{c}
\resizebox{80mm}{!}{\includegraphics{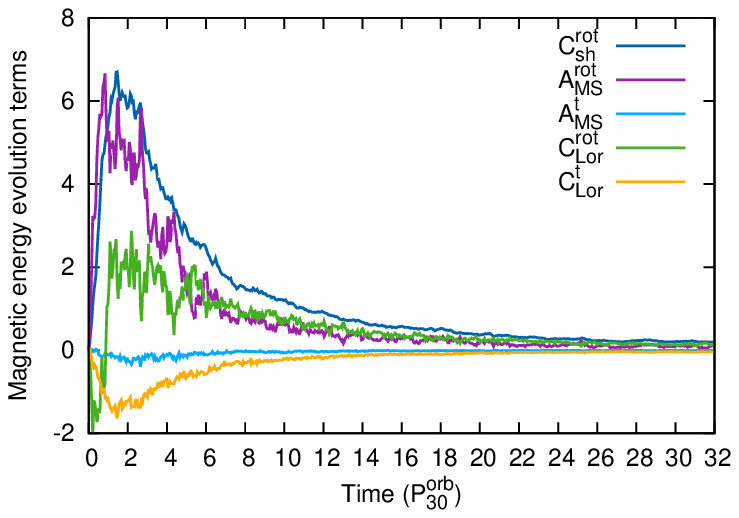}} \\
\resizebox{80mm}{!}{\includegraphics{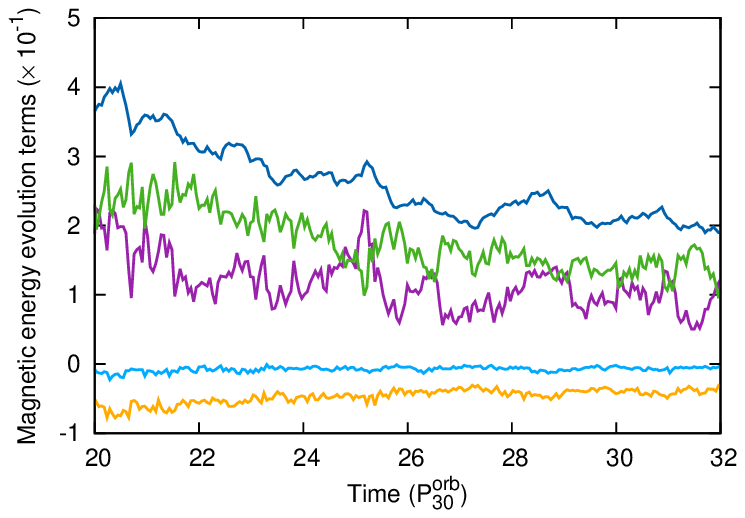}} \\
   \end{tabular}
   \caption{Comparing different terms from the one-zone disk body
     model for model gbl-sr. The top and bottom panels show results
     over the time intervals 0-32 and $20-32~P_{30}^{\rm orb}$,
     respectively. Rates of change of energy are plotted in units of
     $1/P_{30}^{\rm orb}$ - to convert to code units divide the values
     by $P_{30}^{\rm orb}=964$. Note the difference in scale between
     the plots.}
    \label{fig:control_vol2}
  \end{center}
\end{figure}

In Fig.~\ref{fig:control_vol2} we show the result of applying
Eq~(\ref{eqn:mag_energy3}) to model gbl-sr. Magnetic energy
production, which was shown to be predominantly due to $C^{\rm
  rot}_{\rm sh}$ in Fig.~\ref{fig:control_vol}, can now be attributed
to the rates of work done by the mean disk rotation in combination
with Maxwell stresses applied to the boundaries of the volume, $A^{\rm
  rot}_{\rm MS}$, and Lorentz force acting within the volume, $C^{\rm
  rot}_{\rm Lor}$. In contrast, the turbulent velocity field acts to
remove energy from the control volume, as shown by the terms $A^{\rm
  t}_{\rm MS}$ and $C^{\rm t}_{\rm Lor}$. Examining $A^{\rm rot}_{\rm
  MS}$, which involves an integration over the surfaces of the control
volume, one finds the magnitude of the radial surface terms to be much
greater than from the vertical surfaces and, in particular, the inner
radial surface dominates. Therefore, the rate of magnetic energy
production is to a large extent due to the difference between the
rates of work done on the {\it radial} surfaces of the control volume
by Maxwell stresses, and by Lorentz forces within the volume.

\begin{figure}
  \begin{center}
    \begin{tabular}{c}
\resizebox{80mm}{!}{\includegraphics{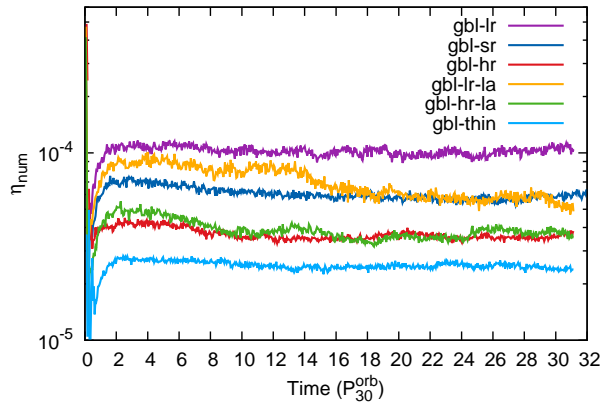}} \\
   \end{tabular}
   \caption{Numerical resistivity, $\eta_{\rm num}$, computed from the
     control volume analysis (see \S~\ref{subsec:mag_energy} and
     \ref{subsec:eta_num}).}
    \label{fig:eta_num}
  \end{center}
\end{figure}

\begin{table}
\begin{center}
  \caption[]{Time averaged numerical resistivity, $\eta_{\rm num}$,
    and magnetic Reynolds number, $Re_{\rm M}$. $\eta_{\rm num}$ is
    estimated using Eq~(\ref{eqn:mag_energy2}) and $Re_{\rm M}=\langle c_{\rm
      s}H \rangle/\eta_{\rm num}$. $\Delta t_{\rm av}$ (second column) is the
    time interval over which time averaging was
    performed.} \label{tab:eta_num}
\begin{tabular}{llll}
  \hline
  Model & $\Delta t_{\rm av}$ & $\overline{\langle\eta_{\rm num}\rangle}$ & $\overline{\langle Re_{\rm M}\rangle}$ \\
  \hline
  gbl-lr & 12-31 & $1.0\times10^{-4}$ & 770 \\
  gbl-sr &  12-31 & $5.8\times10^{-5}$& 1328 \\
  gbl-hr &  12-31 & $3.6\times10^{-5}$ & 2139 \\
  gbl-lr-la & 18-31 & $5.8\times10^{-5}$ & 1328 \\
  gbl-hr-la & 12-31 & $3.7\times10^{-5}$ & 2081 \\
  gbl-thin & 12-31 & $2.5\times10^{-5}$ & 1695 \\
  \hline
\end{tabular}
\end{center}
\end{table}

\subsection{Numerical resistivity}
\label{subsec:eta_num}

Computing the numerical resistivity, $\eta_{\rm num}$, provides
insight into the intrinsic dissipation arising from the simulation
method, embodying the truncated order of accuracy present in commonly
used numerical schemes. For example, in our present investigation we
use third-order accurate spatial reconstruction and second-order
accurate time-stepping. Model gbl-sr returns
$\overline{\langle\eta_{\rm num}\rangle}=6.1\times10^{-5}$ and
$\overline{\langle Re_{\rm M}\rangle}=1273$, whereas, on the basis of
the conclusions drawn by \cite{Fleming:2000}, \cite{Oishi:2011}, and
\cite{Flock:2012b}, sustained turbulence should not be observed for
$Re_{\rm M}\ltsimm 3000$. This disagreement does not appear to be
related to our approximation of a constant $\eta_{\rm num}$ throughout
the control volume, as tests computed for annuli with radial range
$10<r<20$ and $20<r<30$ reveal variations in $\eta_{\rm num}$ of only
$5-10\%$. However, it may be due to our assumption that numerical
resistivity behaves like an Ohmic resistivity. Previous estimates of
numerical resistivity \citep{Fromang:2007a, Simon:2009} adopt a
Fourier analysis of the dissipation term whereby $\eta_{\rm num}$ is
derived from the high wavenumber end of the spectrum. These analyses
reveal that numerical dissipation deviates from Ohmic at low
wavenumbers. Our volume averaged values for $\eta_{\rm num}$ provide
an estimate which is biased towards the large-scales, and thus may be
higher than values at the small scales (i.e. the turbulent dissipation
scale). We do note, however, that the third-order accurate spatial
reconstruction used in our simulations may allow sustained turbulence
at lower $Re_{\rm M}$ than the second-order accuracy used by
\cite{Flock:2012b}. Furthermore, shearing-box boundary conditions
suppress terms in the magnetic energy equation that can supply/sustain
large scale magnetic fields - the importance of global disk boundary
conditions to magnetic field generation is discussed in more detail in
\S~\ref{subsec:bcconv}. Therefore, the large-scale dynamo apparent in
stratified {\it global} disks \citep[Fig.~\ref{fig:meanB} - see
also][]{Arlt:2001, Fromang:2006, O'Neill:2011} may operate effectively
at low $Re_{\rm M}$ \citep[][]{Brandenburg:2009,Kapyla:2011}, meaning
that global disks could exhibit sustained turbulence at lower $Re_{\rm
  M}$ than in a shearing-box \citep{Fleming:2000,Oishi:2011}. For
further discussion of numerical resistivity see \cite{Hirose:2006} and
\cite{Hawley:2011}.

In summary, there are two main reasons for the difference by a factor
$\sim3$ for the critical Reynolds number for the maintenance of
turbulence in global disk models compared to the work of
\cite{Flock:2012b}: (1) Differing mathematical approaches to the
estimation of the resistivity - \cite{Flock:2012b} include an Ohmic
resistive term specifically in their simulations, whereas we estimate
it using an Ohmic model for the numerical resistivity, (2) The
\citeauthor{Flock:2012b} simulations are spatially second order
accurate whereas our simulations (and analysis) are spatially third
order accurate.

Examining the results for models gbl-lr, gbl-sr, and gbl-hr in
Fig.~\ref{fig:eta_num} and Table~\ref{tab:eta_num}, there is the
consistent trend that as the resolution is increased (and the cell
aspect ratio is kept fixed) the value of $\eta_{\rm num}$
decreases. For example, between models gbl-lr and gbl-sr the
resolution has been increased by a factor of 1.5 resulting in a
decrease in $\eta_{\rm num}$ by a factor of 1.8. Based on the results
for these three simulations we find $Re_{\rm M} \approx 0.45 (n_{\rm
  r})^{1.3}$, where $n_{\rm r}$ is the number of cells in the radial
direction, leading to an estimated resolution requirement of $n_{\rm
  r}\times n_{\theta} \times n_{\phi} \simeq 2600\times860 \times
1000\;$cells to achieve a magnetic Reynolds number, $Re_{\rm
  M}=10^4$. This poses a significant computational
challenge\footnote{This may, however, be alleviated using an orbital
  advection/FARGO scheme \citep[e.g.][]{Sorathia:2012,
    Mignone:2012}.}. Note that our derived scaling for numerical
resistivity is identical to that found by \cite{Simon:2009} for
unstratified {\it net flux} shearing-boxes.

Somewhat surprisingly, model gbl-lr-la displays a lower value for
$\eta_{\rm num}$, and thus higher $Re_{\rm M}$, than gbl-lr despite
the former having a larger cell aspect ratio. Considering the lower
level of turbulent activity in model gbl-lr-la compared to gbl-lr (see
Figs.~\ref{fig:alpha} and \ref{fig:curlB2}), this shows that numerical
resistivity scales with the turbulent motion of the magnetic field,
i.e. a larger value of $|j|$ causes a larger net truncation error. In
model gbl-lr-la, turbulent activity wanes for $t\gtsimm 14~P^{\rm
  orb}_{30}$, and simultaneously the value of $\eta_{\rm num}$ dips
(Fig.~\ref{fig:eta_num}).

\section{Boundary conditions and convergence}
\label{sec:convergence}

The question of convergence in simulation studies of magnetized
accretion disk turbulence has been long-standing
\citep[e.g.][]{Hawley:1995,Stone:1996, Sano:2004, Fromang:2007a,
  Simon:2009, Guan:2009, Johansen:2009, Davis:2010, Hawley:2011,
  Sorathia:2012}. It is clear that the development, or
initial presence, of large-scale magnetic field components is a vital
ingredient in enabling convergence with increasing simulation
resolution - see the discussion in the second paragraph of \S~\ref{sec:control_volume}. When
present, large-scale magnetic fields can replenish low wavenumber
magnetic energy. This is a pre-requisite for convergence since
otherwise the reservoir of magnetic energy on the largest scales is
drained by the turbulent cascading of magnetic energy to smaller scales.
We show in this section that the simulation boundary conditions
dictate whether large scale mean fields can grow and thereby promote
convergence. In doing so, we consider three different classes of
simulation: (1) Unstratified shearing boxes; (2) Stratified shearing
boxes; (3) Global, stratified disks and (4) Global, unstratified
disks.  In the simplest case - the unstratified shearing-box with
periodic boundary conditions - mean radial and vertical fields cannot
readily evolve. When stratification is introduced the associated
interface between the disk body and corona, relaxes the constraint on
mean radial field growth such that an $\alpha-\Omega$ dynamo can
operate effectively.  In global models, mean fields grow relatively
quickly, enabling large-scale dynamo activity and magnetic energy
replenishment. A key result of this analysis is that lower simulation
resolution is required in stratified global models compared to
shearing-boxes because the large-scale radial gradients enabled by
open radial boundaries permit a larger magnitude contribution to the
creation of magnetic energy from the Lorentz force terms. Thus,
convergence is attained at lower resolutions in global models than in
shearing-boxes.

\subsection{The magnetic energy balance in accretion disk turbulence}
\label{subsec:general_evolution}

Irrespective of the specific setup (unstratified/stratified shearing
box, global simulation) numerical simulations of magnetorotational
turbulence exhibit common features. During the early phases of
simulation evolution, the MRI develops and the subsequent magnetic
field amplification causes a sharp rise in the magnetic energy,
$\langle u_{\rm B}\rangle$. Magnetic energy built-up during the
initial transient growth phase supports optimal MRI growth and
turbulent driving, which in-turn dissipates magnetic energy via the
resistivity and the current density, $j_{\rm i}=[\nabla \times B]_{\rm
  i}$. Magnetic energy subsides, and a state is approached where
magnetic field production and turbulent dissipation come into
balance. This latter stage is the QSS.

Informed by the analysis in \S~\ref{sec:control_volume}, we write the
steady-state magnetic energy evolution equation as (see
Eq~(\ref{eqn:mag_energy3}):
\begin{equation}
   \dot{U}_{\rm B} = C_{\rm Lor} + A_{\rm MS} + D_{\rm num} \approx 0, \label{eqn:mag_bal}
\end{equation}
where the separate Lorentz and magnetic stress terms for rotational and turbulent contributions
are combined in the following terms:
\begin{eqnarray}
C_{\rm Lor} &=&  - \int v_{\rm i} \frac{\partial M^{\rm B}_{\rm
      ij}}{\partial x_{\rm j}} dV = - \int v_{\rm i} F^{\rm L}_{\rm i} dV,\\
A_{\rm MS} &=& \oint M^{\rm B}_{\rm ij} v_{\rm j} n_{\rm i}
  dS, 
\end{eqnarray}
where $v_{\rm i}$ is the total (rotational plus turbulent) velocity,
$M^{\rm B}_{\rm ij}$ is the Maxwell stress tensor and $F^{\rm L}_{\rm
  i}$ the Lorentz force. From \S~\ref{subsec:mag_energy} we have
$D_{\rm num}\approx -\eta_{\rm num}\int |j|^2 dV$, so that
Eq~(\ref{eqn:mag_bal}) for the magnetic energy balance now reads:
\begin{eqnarray}
 \oint v_{\rm i} M^{\rm B}_{\rm ij} n_{\rm j} dS
- \int v_{\rm i}\frac{\partial M_{\rm ij}^{\rm B}}{\partial x_{\rm
  j}} \> dV 
  \approx \eta_{\rm num} \int |j|^2 dV. \label{eqn:balance}
\end{eqnarray}
Eq~(\ref{eqn:balance}) states that the achievement of a QSS requires the rate
of work done on the surfaces of the control volume by Maxwell
stresses, and by Lorentz forces within the volume, to be balanced by
dissipation. Magnetic field saturation and the quasi-steady state are
solutions to Eq~(\ref{eqn:balance}).

\subsection{Integrated form of the induction equation}

In our analysis below, the interplay between the induction equation 
and the boundary conditions also plays an important role. We begin with the induction
equation,
\begin{equation}
  \frac{\partial B_{\rm i}}{\partial t} = \frac{\partial}{\partial x_{\rm j}} (v_{\rm i}B_{\rm j} 
- B_{\rm j} v_{\rm i}) + \eta_{\rm num} \nabla^2 B_{\rm i}, \label{eqn:induction3}
\end{equation}
where a numerical resistive term is included for consistency with the
magnetic energy balance Eq~(\ref{eqn:balance}).  Integrating
Eq~(\ref{eqn:induction3}) over a control volume, $V$, with bounding
surface $S$, we have:
\begin{equation}
\frac{\partial}{\partial t} \int B_{\rm i} dV = \oint 
\left( v_{\rm i}B_{\rm j} - B_{\rm i}v_{\rm j}    
+ \eta_{\rm num} \frac{\partial B_{\rm i}}{\partial x_{\rm j}} \right)
\> dS_{\rm j}. \label{eqn:induction4}
\end{equation}
where $dS_j$ is the element of surface area and we have assumed $\eta_{\rm num}$ to be approximately spatially constant.

The surface $S$ bounding the control volume as well as the boundary
conditions on $S$ take several different forms depending upon the
simulation - stratified/unstratified shearing box,
stratified/unstratified global disk. However, in general we can use
the coordinate convention introduced for shearing boxes
\citep{Hawley:1995}, adapting the mathematical analysis in each of the
different cases. Therefore, the coordinates $x$, $y$ and $z$
correspond to the radial, azimuthal, and vertical directions in the
control volume, respectively. The surface integral then involves three
separate integrals over the $x$, $y$ and $z$ faces, which we denote by
$x=x_1, \> x_2$, $y=y_1, \> y_2$ and $z= z_1, \> z_2$ respectively. In
our representation of the integrated induction equation we introduce
the resistive flux,
\begin{equation}
F_i^{\rm res} = \oint_S \eta_{\rm num} \frac {\partial B_{\rm
    i}}{\partial x_{\rm j}} dS_{\rm j}.
\end{equation}
This represents a diffusion of magnetic field, resulting from resistive effects through the bounding surface $S$.
The integrated induction equation becomes for each coordinate:
\begin{eqnarray}
\frac{\partial}{\partial t} \int B_{\rm x} dV &=& F^{\rm res}_x + 
\int_{y_2} \left( v_x B_y - v_y B_x \right) \> dS_y \nonumber \\
&& -\int_{y_1} \left( v_x B_y - v_y B_x \right) \> dS_y \nonumber \\
&& + \int_{z_2} \left( v_x B_z - v_z B_x  \right) \> dS_z \nonumber \\
&& - \int_{z_1} \left( v_x B_z - v_z B_x  \right) \> dS_{z}, \label{induction_x}\\
\frac{\partial}{\partial t} \int B_{\rm y} dV &=& F^{\rm res}_y +
\int_{x_2} \left( v_y B_x - v_x B_y  \right) \> dS_x \nonumber \\
&& -\int_{x_1} \left( v_y B_x - v_x B_y  \right) \> dS_x \nonumber \\
&& + \int_{z_2} \left( v_y B_z - v_z B_y  \right) \> dS_z \nonumber \\
&& - \int_{z_1} \left( v_y B_z - v_z B_y  \right)  \> dS_{z}, \label{induction_y}\\
\frac{\partial}{\partial t} \int B_{\rm z} dV &=& F^{\rm res}_z +
\int_{x_2} \left( v_z B_x - v_x B_z \right) \> dS_x \nonumber \\
&& -\int_{x_1} \left( v_z B_x - v_x B_z  \right) \> dS_x \nonumber \\
&& + \int_{y_2} \left( v_z B_y - v_y B_z \right) \> dS_y \nonumber \\
&& - \int_{y_1} \left( v_z B_y - v_y B_z \right)  \> dS_{y}. \label{induction_z}
\end{eqnarray}
The surface integrals in Eqs~(\ref{induction_x}) - (\ref{induction_z})
show the influence of the velocity and magnetic field values at the
boundaries on the volume integrated field within the control volume.

\subsection{Dependence of convergence on boundary conditions and
  magnetic field configuration}
\label{subsec:bcconv}

In the following sections we utilise our description of the magnetic
energy balance combined with inferences from the induction equation to
describe how the convergence properties of simulations with different
numerical setups can be readily understood in terms of the respective
boundary conditions and net magnetic field configuration.

\subsubsection{Unstratified shearing-box}
\label{subsubsec:sb}

In this case the model is a periodic box with background shear applied
via source terms in the momentum equation. The shearing-box method is
used to represent a small patch of an accretion disk in a Cartesian
coordinate system such that $x$, $y$, and $z$ correspond to the
radial, azimuthal, and vertical directions, respectively; the
corresponding lengths of each side of the box are $L_{\rm x}$, $L_{\rm
  y}$ and $L_{\rm z}$ \citep[see][for further
details]{Hawley:1995}. The boundaries of the control volume in this
setup are the boundaries of the computational domain. For an
unstratified shearing-box, the following shearing-periodic boundary
conditions are applied in the radial ($x$), azimuthal ($y$) and
vertical ($z$) directions for all dynamical variables $f(x,y, z)$
except the azimuthal velocity. Let $q = d \ln \Omega /d \ln R$ be the
shear parameter (=3/2 for a Keplerian disk), then the $x-$, $y-$ and
$z-$boundary conditions are:
\begin{eqnarray}
f(x+L_{\rm x},y,z) &=& f(x,y+q\Omega L_{\rm x}t,z), \label{spbc_x}\\
f(x,y+L_{\rm y},z) &=& f(x,y,z), \label{spbc_y}\\
f(x,y,z+L_{\rm z}) &=& f(x,y,z). \label{spbc_z}
\end{eqnarray}
The exception is the azimuthal velocity, which satisfies the 
above $y$- and $z$- boundary
conditions but whose $x$-boundary condition is:
\begin{equation}
  v_{\rm y} (x+L_x, y, z) = v_{\rm y} (x,y+q \Omega L_{\rm x} t, z) + q
  \Omega L_{\rm x}.
\label{spbc_vy}
\end{equation}
Applying these boundary conditions to
Eq~(\ref{eqn:balance}), and noting that $v_{\rm i}B_{\rm i} \approx
B_{\rm y}v_{\rm y}$, we have,
\begin{equation}
  -q \Omega L_{\rm x} \int_{\rm x_1} B_{\rm x}B_{\rm y} dS_{\rm x} +
  C_{\rm Lor} \approx \eta_{\rm num}
  \int |j|^2 dV, \label{eqn:sb}
\end{equation}
where, as noted, $x_1$ refers to the inner radial boundary and
$dS_{\rm x}$ is the corresponding element of surface area. Furthermore, noting
that structures are typically elongated in the $y$-direction, then
$\partial/\partial y \ll \partial /\partial x$ or $\partial/\partial
z$, and retaining the largest terms (those linear in $v_{\rm y}$ or
$B_{\rm y}$), one finds,
\begin{eqnarray}
  -q \Omega L_{\rm x} \int_{\rm x_1} B_{\rm x}B_{\rm y} dS_{\rm x} -
  \int v_{\rm y} B_{\rm x}\frac{\partial B_{\rm y}}{\partial x} dV
  - \int v_{\rm y} B_{\rm z}\frac{\partial B_{\rm y}}{\partial z} dV
 \nonumber\\
 \approx \eta_{\rm num} \int |j|^2 dV. \label{eqn:sb2}
\end{eqnarray}
The crucial feature of Eq~(\ref{eqn:sb2}) is that magnetic energy is
produced by a combination of the $x-y$ component of the Maxwell stress
at the radial boundary and Lorentz forces doing work within the volume;
the Lorentz forces depend on the radial and vertical field components 
as well as the 
radial and vertical gradient in $B_{\rm y}$. However, the contributions 
from the second
and third terms on the LHS of Eq~(\ref{eqn:sb2}) are negligible on
large scales if there is zero-net radial and vertical magnetic field,
and/or no radial or vertical gradient in $B_{\rm y}$. 

We now consider the implications of the induction equation for the
large scale radial and vertical magnetic fields. 
Inserting the shearing-periodic boundary conditions (\ref{spbc_x}) -- (\ref{spbc_vy}) into the 
integrated induction Eqs~(\ref{induction_x}) -- (\ref{induction_z}), we obtain:
\begin{eqnarray}
  \frac{\partial}{\partial t}\int B_{\rm x} dV &=& F^{\rm res}_{\rm x}, \label{bx_vol}\\
  \frac{\partial}{\partial t}\int B_{\rm y} dV &=& F^{\rm res}_y -q \Omega L_{\rm x}
  \int_{\rm in} B_{\rm x} dS_{\rm x}, 
  \label{by_vol}\\
  \frac{\partial}{\partial t}\int B_{\rm z} dV &=& F^{\rm res}_{\rm z}.  \label{bz_vol}
\end{eqnarray}
Eq~(\ref{by_vol}) for the azimuthal field shows that it evolves as a
result of resistive diffusion but also, and more importantly as a
result of the combined action of velocity shear and the radial field.
However, Eq~(\ref{bx_vol}) for the integrated radial field and
Eq~(\ref{bz_vol}) for the vertical field show that these components
evolve solely under the action of resistive diffusion and there is no
influence from the boundary values of the velocity combined with
existing field components. If the net fluxes associated with $B_x$ or
$B_z$ are initially zero and start to build up within the volume then
the diffusion terms will act to dissipate these fluxes and they will
remain at near zero levels. Hence, initially zero net radial and
vertical fields do not develop significant components on the largest
scale in the box. On the other hand a net flux vertical field prevails
on the timescale of the simulation, and will maintain a component on
the largest realizable scale in the simulation domain. This is
notwithstanding the effect of diffusion since maintaining a non-zero
boundary value of $B_z$ minimises diffusion of $B_z$ out of the volume
as a result of the gradient in $B_z$ being close to zero.

Since magnetorotational turbulence extracts energy from the	
largest scales and drives it towards the smallest scales
\citep{Fromang:2007a,Johansen:2009, Lesur:2011}, the preservation of a 
net vertical field
fixes the injection of magnetic energy at the scale of the box, thus
replenishing the low wavenumber end of the magnetic energy power
spectrum. In contrast, in a zero-net flux, unstratified shearing-box,
the finite reservoir of magnetic energy at the low wavenumber end of
the scale is depleted by turbulent driving. 

Achieving convergence is, therefore, related to the presence of
magnetic energy injection by Lorentz forces on the largest realisable
scales and correspondingly the existence of large scale vertical
and/or radial field on those scales.  Our analysis explains the
results in \citet{Simon:2009} who compared zero-net flux and net flux
simulations. They demonstrated that energy injection - represented by
the Fourier space analogue of our shear term, $C_{\rm sh}^{\rm rot}$ -
continues to rise as one tends towards the largest scales in the box
in the case of net flux simulations, whereas it plateaus for the zero
net flux simulations. This indicates that in evolved zero-net flux
turbulence (in an unstratified shearing-box), magnetic energy is not
replenished effectively on the largest scales, and this is also
consistent with the lack of a large-scale dynamo \citep{Vishniac:2009,
  Bodo:2011, Kapyla:2011}.

The above analysis also relates to another well-studied 
problem within the literature, namely the origin of the lack of 
convergence with increasing resolution in
unstratified, zero-net flux shearing-box simulations 
\citep[e.g.][]{Fromang:2007a,
  Pessah:2007, Regev:2008, Vishniac:2009, Kapyla:2011,
  Bodo:2011}. As we have shown above, 
unstratified, zero-net flux shearing-box simulations with periodic
boundary conditions render the Lorentz force term ineffective at
injecting magnetic energy, and thus a large-scale mean field cannot 
develop. Hence, when a QSS with turbulent
transport of magnetic energy from larger to smaller scales
establishes, it must suffice with the largest scale field
available: A small-scale dynamo operates, for which the stress scales
proportionately to the resistivity
\citep{Vishniac:2009,Bodo:2011}. Commencing the simulation with a net
radial/vertical field \citep{Hawley:1995, Sano:2004, Simon:2009,
  Guan:2009}, or adopting alternative boundary conditions which permit
the development of mean fields within a few orbital periods
\citep[e.g. vertical field boundary conditions][]{Kapyla:2011},
enables convergence. Our above analysis provides insight into why
these strategies are successful.

\subsubsection{Stratified shearing box:} 

We now consider \emph{stratified} shearing box simulations in which
the vertical component of gravity is included. In the context of the
current analysis, the results of zero-net flux simulations by
\cite{Davis:2010} and \cite{Oishi:2011} are useful as periodicity is
applied at the boundaries of the computational domain (including the
vertical boundary at $|z|=2H$), in common with the unstratified
simulations discussed in \S~\ref{subsubsec:sb}. However, unlike the
zero-net flux unstratified shearing-boxes described above, the
\cite{Davis:2010} models converge with increasing resolution. The
crucial difference is that stratification provides a means for the
disk to repartition magnetic flux so that the disk body can overcome
the magnetic flux constraint and generate large scale magnetic fields.
In essence, stratification introduces an internal open boundary
between the disk body and the coronal region. From the results
presented in \cite{Davis:2010}, we infer this open boundary to lie at
$|z_{1,2}|\simeq 1 -1.5~H$, which we adopt in the following analysis.
With the boundaries $z=z_1$ and $z=z_2$ now not constrained to be
periodic the integrated induction equations are:
\begin{eqnarray}
  \frac{\partial}{\partial t}\int B_{\rm x} dV &=&
  F_x^{\rm res} +
  \int_{z_2} v_{\rm x}B_{\rm z} \> dS_{\rm z}  
  - \int_{z_1} v_{\rm x}B_{\rm z} \> dS_{\rm z} \nonumber\\
  && - \int_{z_2} B_{\rm x}v_{\rm z} \> dS_{\rm z}  
  + \int_{z_1} B_{\rm x}v_{\rm z} \> dS_{\rm z}, \label{ssb:bx}\\
  \frac{\partial}{\partial t}\int B_{\rm y} dV &=& F_y^{\rm res} 
  -q \Omega L_{\rm x}
  \int_{\rm x_1} B_{\rm x} \> dS_{\rm x} \nonumber \\
  && + \int_{z_2} v_{\rm y}B_{\rm z}  \> dS_{\rm z} - \int_{z_1}
  v_{\rm y}B_{\rm z} \> dS_{\rm z} \nonumber\\
  && - \int_{z_2} B_{\rm y}v_{\rm z} \> dS_{\rm z}  
  + \int_{z_1} B_{\rm y}v_{\rm z} \> dS_{\rm z}, 
  \label{ssb:by}\\
  \frac{\partial}{\partial t}\int B_{\rm z} dV &=& F_z^{\rm res}. \label{ssb:bz}
\end{eqnarray}
The non-periodic boundary conditions on the faces $z=z_1$ and $z_2$
introduce additional driving terms into the radial equation involving
the terms $v_{\rm x} B_{\rm z}$ and $B_{\rm x} v_{\rm z}$. Given the
zero-net flux condition, these terms are important if the fluctuations
in $v_{\rm x}$ and $B_{\rm z}$ or $B_{\rm x}$ and $v_{\rm z}$ are
correlated. If this is the case, then Eq~(\ref{ssb:bx}) opens up the
possibility of net radial field development and an $\alpha-\Omega$
dynamo.

Considering the magnetic energy equation, we apply periodic boundaries in the azimuthal and radial directions
and an open vertical boundary condition to Eq~(\ref{eqn:balance}), and
retain only the dominant terms, to obtain:
\begin{eqnarray}
  -q \Omega L_{\rm x} \int_{\rm x_1} B_{\rm x}B_{\rm y} \> dS_{\rm x} +
  \int_{z_2} v_{\rm y}B_{\rm y}B_{\rm z} \> dS_{\rm z} - \int_{z_1} v_{\rm y}B_{\rm y}B_{\rm z} \> dS_{\rm z}
  \nonumber \\
  - \int v_{\rm y} B_{\rm z}\frac{\partial B_{\rm y}}{\partial z} dV 
  - \int v_{\rm y} B_{\rm x}\frac{\partial B_{\rm y}}{\partial x} dV
  \approx \eta_{\rm num} \int |j|^2 dV. \label{eqn:ssb}
\end{eqnarray}
where the additional term compared to the energy equation for an
unstratified disk, Eq~(\ref{eqn:sb2}), arises from the work done on
the disk-corona vertical boundary by Maxwell stresses. Despite the
presence of the disk-corona interface, a large-scale vertical magnetic
field does not develop (see Eq~\ref{ssb:bz}). Thus the second, third
and fourth terms on the LHS of Eq~(\ref{eqn:ssb}) are negligible. We
are left with:
\begin{eqnarray}
  -q \Omega L_{\rm x} \int_{\rm x_1} B_{\rm x}B_{\rm y} dS_{\rm x} 
  - \int v_{\rm y} B_{\rm x}\frac{\partial B_{\rm y}}{\partial x} dV
  \approx \eta_{\rm num} \int |j|^2 dV. \label{eqn:ssb2}
\end{eqnarray}
We conjecture that the \citeauthor{Davis:2010} simulations converge
due to the terms involving $B_{\rm x}$ and $B_{\rm y}$ on the LHS of
Eq~(\ref{eqn:ssb2}), where the introduction of stratification permits
the development of a large scale radial magnetic field which
combines with the azimuthal field to enable an $\alpha-\Omega$ dynamo to
operate. As the simulation resolution is increased, the resolution of
MRI modes improves. At a critical resolution, the most unstable
wavelength becomes resolved and a further increase in resolution
ceases to provide additional MRI growth \citep[because wavelengths
shorter than the most unstable mode are stable -][]{BH92,BH98}. The
contribution to the power input from Maxwell stresses and Lorentz
forces (the first and second terms on the LHS of Eq~\ref{eqn:ssb2})
asymptote towards constant values as radial and azimuthal MRI mode
growth converges. We note that the above argument is consistent with
the results of \cite{Oishi:2011}, as the presence of a disk-corona
interface relaxes the helicity conservation constraint for dynamo
quenching.

\subsubsection{Global stratified disk simulation}
\label{subsubsec:global}

In global stratified disk models, such as the ones we have described
in this paper, periodic boundary conditions are applied in the
azimuthal direction and the radial and vertical boundaries of the disk
body are open. Assuming symmetry about the mid-plane, one may take the
vertical surfaces to be anti-periodic. The control volume in our
global stratified disk simulations is, in spherical polars, $10 \leq r
\leq 30$, $ \theta_{2H/R} <\theta - \pi/2 < \theta_{2H/R}$,
where $\theta_{2H/R}=\tan^{-1}(2\langle H/R\rangle)$, and $0 \leq \phi
\leq \pi/2$.  The magnetic energy equation for this control volume,
which follows from applying appropriate boundary conditions to the
magnetic energy Eq~(\ref{eqn:balance}) and retaining the dominant
terms, is:
\begin{eqnarray}
  \int_{r_2} B_{\rm r}B_{\phi }v_{\phi} dS_{\rm r} 
  - \int_{r_1}   B_{\rm r}B_{\phi }v_{\phi} dS_{\rm r}  
  - 2 \int_{\theta_1} B_{\theta}B_{\phi}v_{\phi} dS_{\theta} \nonumber \\
  + \> C_{\rm Lor} = \eta_{\rm num}
  \int |j|^2 dV, \label{eqn:global}
\end{eqnarray}
where the element of volume, $dV = r^2 \sin\theta dr d\theta d\phi$,
the surface element orthogonal to $\theta=\rm constant$ is $dS_\theta
= r \sin \theta dr d\phi$, and the surface element orthogonal to
$r=\rm constant$ is $dS_{\rm r}=r^2 d\theta d\phi$. As noted in the last
paragraph of \S~\ref{subsec:mag_energy}, the rate of work done by
Maxwell stresses on the inner radial boundary is far greater than that
done on the outer radial and vertical ($\theta$-direction) boundaries,
so that the second term on the LHS of Eq~(\ref{eqn:global}) dominates
the first and third terms. Also, the largest contributor to $C_{\rm
  Lor}$ is $\int v_{\phi}B_{\rm r}(\partial B_{\phi}/\partial r)
dV$. Hence,
\begin{eqnarray}
  -\int_{r_1} B_{\rm r}B_{\phi }v_{\phi} dS_{\rm r} - \int v_{\phi}B_{\rm r}\frac{\partial
    B_{\phi}}{\partial r} dV \simeq \eta_{\rm num} \int |j|^2 dV. \label{eqn:global2}
\end{eqnarray}
Note the similarity of the above equation with that governing the
stratified shearing-box simulations (Eq~\ref{eqn:ssb2}). The distinct
difference is in the magnitude of the terms on the LHS, which depend
on $B_{\rm r}$, $B_{\phi}$, and the radial gradient of
$B_{\phi}$. Similar to the stratified shearing-box, for magnetic
energy production and dissipation to converge, the growth of $B_{\rm
  r}$ and $B_{\phi}$ must be well resolved. In contrast, however,
because of the large-scale radial gradient in $B_{\phi}$ which is
present in a global model \citep[see, e.g., figure 12
of][]{Flock:2011} but which is suppressed by periodic radial boundary
conditions in a shearing-box, the second term is larger in a global
model. This term is appreciable in magnitude for a global disk with
open radial and vertical boundaries as a result of the development of
a significant net radial field. We can indicate how this arises by
considering the integrated induction equation for the radial field,
this time in spherical polar coordinates, with periodic azimuthal
boundary conditions, {\em viz.},
\begin{eqnarray}
  \frac{\partial}{\partial t} \int B_r \> dV &=& 
  \int_{\theta_2} (v_r B_\theta - v_\theta B_r) \> dS_\theta \nonumber \\
  &&- \int_{\theta_1} (v_r B_\theta - v_\theta B_r) \> dS_\theta \nonumber \\
  && \approx - 2 \int_{\theta_1} (v_r B_\theta - v_\theta B_r) \> dS_{\theta}.
\label{induction_brsph}
\end{eqnarray}
In this equation we have omitted the resistive diffusive terms in
order to concentrate on the driving terms for $B_r$ and we have made
use of the approximate anti-symmetry of the $\theta=\theta_1$ and
$\theta=\theta_2$ surfaces.

Returning to the implications for the magnetic energy equation, we
note that since periodicity is not
applied to Maxwell stresses doing work on the radial boundaries in a
global model, they contribute more power. Therefore, in a global
model, Lorentz forces within the volume and Maxwell stresses at the
boundaries of the disk inject more power at a lower resolutions than
they do in a shearing-box. We emphasise that this is the result of
radially periodic boundary conditions.  Thus, keeping the cell aspect
ratio constant and progressively increasing simulation resolution as
we have done with models gbl-lr, gbl-sr, and gbl-hr, and as
\cite{Davis:2010} did with their stratified shearing-box simulations,
one should achieve apparent convergence in $\langle\alpha_{\rm
  P}\rangle$ at lower simulation resolutions (i.e.  cells/$H$ in the
vertical direction) for global disk simulations than for stratified
shearing-box simulations; \cite{Davis:2010} found convergence at
64-128 cells/$H$ in the vertical direction, whereas we find
convergence at $\sim 27$ cells/$H$ in the vertical direction (see
Tables~\ref{tab:model_list} and \ref{tab:global_models}).

Comparing the magnitude of the surface integral terms on the LHS of
Eqs~(\ref{eqn:ssb2}) and (\ref{eqn:global2}), for the magnetic energy
of the unstratified shearing-box and global stratified disk
respectively, we find that $|\int_{r_1} B_{\rm r}B_{\phi}v_{\phi}
dS_{\rm r}|$ is greater than $|q\Omega L_{\rm x}\int_{\rm x_1}B_{\rm
  x}B_{\rm y}dS_{\rm x}|$ by a factor of a few tens. Hence, a global
disk more readily supports a high $\langle\alpha_{\rm P}\rangle$
because the radial boundary condition allows more power to be
delivered to the disk body to counteract the removal of energy by
turbulent dissipation ($\eta_{\rm num}\int |j|^2 dV$). Stating this in
a more general context, periodic boundary conditions on the magnetic
field prevent the establishment of large-scale gradients in the
Maxwell stresses, restricting the power that can be delivered to the
disk by Lorentz forces and surface stresses.

We note, however, that in the absence of an explicit resistivity,
simulations performed at different resolutions will, at some late
time, diverge. This is because in the quasi-steady state the disk is
continuing to evolve on the (slow) resistive timescale. If one relies
on numerical resistivity, this timescale is dictated by 
$\eta_{\rm num}$. We anticipate that as global disk simulations integrated over
many hundreds of orbits become more feasible, this result will be
realised. In fact, the results of \cite{Sorathia:2012} already show
this for unstratified global disk simulations.

\subsubsection{Global unstratified disk simulation}
\label{subsubsec:global_unstrat}

We complete this analysis by considering the global
\emph{unstratified} disk models presented by \cite{Sorathia:2012}
showing what differences the vertical periodic boundary conditions
make in that case. We have presented in Eqs~(\ref{induction_x}) -
(\ref{induction_z}) the volume-integrated induction equations
appropriate for Cartesian shearing boxes. \citet{Sorathia:2012} employ
cylindrical polar coordinates $R,\phi,z$ so that we present the
following induction equations in that coordinate system. We are
interested in the driving terms for these components so that we omit
the diffusive terms in these equations, which have a complex form and
do not add anything to the discussion.

The volume-integrated equations, assuming periodicity in the azimuthal direction are:
\begin{eqnarray}
\frac{\partial}{\partial t} \int B_R \> dV &=& 
\int_{z_2} (v_R B_z - v_z B_R) \> dS_z \nonumber \\
&& - \int_{z_1} (v_R B_z - v_z B_R) dS_{\rm z}, \label{bRcyl_vol}\\
\frac{\partial}{\partial t} \int \frac{B_\phi}{R} \> dV &=&
\int_{z_2} (\frac {v_\phi}{R} B_z  - v_z \frac{B_\phi}{R}) \> dS_z \nonumber \\
&& - \int_{z_1} \left(\frac {v_\phi}{R} B_z  - v_z \frac{B_\phi}{R} \right) \> dS_z \nonumber \\
&& + \int_{R_2} \left(\frac {v_\phi}{R} B_R - v_R \frac{B_\phi}{R} \right) \> dS_R \nonumber \\
&& -  \int_{R_1} \left( \frac {v_\phi}{R} B_R - v_R \frac{B_\phi}{R}
\right) \> dS_{\rm R}, \\
\frac{\partial}{\partial t} \int B_z \> dV &=& 
\int_{R_2} (v_z B_R - v_R B_z) \> dS_R \nonumber \\
&& - \int_{R_1} (v_z B_R - v_R B_z) \> dS_{\rm R}.
\label{bzcyl_vol}
\end{eqnarray}
where the element of volume $dV= R dR d\phi dz$ and the respective
elements of area are $dS_R = R d\phi dz$ (orthogonal to $R=\rm
constant$) and $dS_z = R dR d\phi$ (orthogonal to $z=\rm constant$).
\citet{Sorathia:2012} use open radial boundaries and periodic boundary
conditions in the vertical and azimuthal
direction. \citet{Sorathia:2012} use open radial boundaries and
periodic boundary conditions in the vertical and azimuthal
direction. Since the radial boundaries are open, Eq~(\ref{bzcyl_vol})
shows that a mean vertical field can grow irrespective of whether it
is initially zero. However, with a periodic $z$-boundary, the RHS of
Eq~(\ref{bRcyl_vol}) is identically zero and a large scale radial
field cannot develop.

Hence, in this case, the magnetic energy equation with only dominant
terms retained reads, 
\begin{eqnarray}
  -\int_{R_1} B_{\rm R} B_{\phi }v_{\phi} dS_{\rm R} 
  - \int v_{\phi}B_{\rm R}\frac{\partial
    B_{\phi}}{\partial R} dV - \int v_{\phi}B_{\rm z}\frac{\partial
    B_{\phi}}{\partial z} dV \nonumber\\
  \simeq \eta_{\rm num} \int |j|^2 dV, \label{eqn:global3}
\end{eqnarray}
where we have adopted cylindrical coordinates ($R,\phi,z$) for
consistency with the work of \cite{Sorathia:2012}. The second and third
terms in this equation are likely to be small compared to the first 
(and will contribute little
to maintaining magnetic power on the largest scales), because there is
no large-scale radial field, and because the disk is unstratified so that
there is no appreciable vertical gradient in $B_{\phi}$. Hence, 
Eq~(\ref{eqn:global3}) simplifies to,
\begin{eqnarray}
-\int_{\rm R_1} B_{\rm R}B_{\phi }v_{\phi} dS_{\rm R}  \simeq \eta_{\rm num} \int |j|^2 dV, \label{eqn:global3}
\end{eqnarray}
where the remaining Maxwell stress term provides the power input on
the largest scales. Note that this term does not require a large-scale
net/mean radial field, and has a considerable magnitude purely due to
open radial boundary conditions causing a contrast in surface
integrals at opposing boundaries. The apparent convergence in
$\langle\alpha_{\rm P}\rangle$ present in the results of
\cite{Sorathia:2012} therefore hinges on adequate resolution of the
radial and azimuthal magnetic field. Hence, one may anticipate that
stratified and unstratified global models will converge at the same
resolution. This is apparent from a comparison of our results with
those of \cite{Sorathia:2012}.

It is noteworthy that although stratified and unstratified global
models do appear to converge at similar resolutions, this may be
facilitated by different mechanisms in each case. Essentially, because
a mean radial field cannot develop in an unstratified global model,
maintenance of large scale magnetic energy is not facilitated by a
large-scale $\alpha-\Omega$ dynamo, but must be aided by some other
mechanism - see, for example, \cite{Lesur:2008}. Hence, by
construction, periodic vertical boundary conditions place more demand
on the azimuthal and vertical fields to sustain turbulent energy on
the largest scales. Considering that astrophysical disks are
stratified, this seems an unrealistic approximation to a real disk.

\begin{figure}
  \begin{center}
    \begin{tabular}{c}
    \resizebox{80mm}{!}{\includegraphics{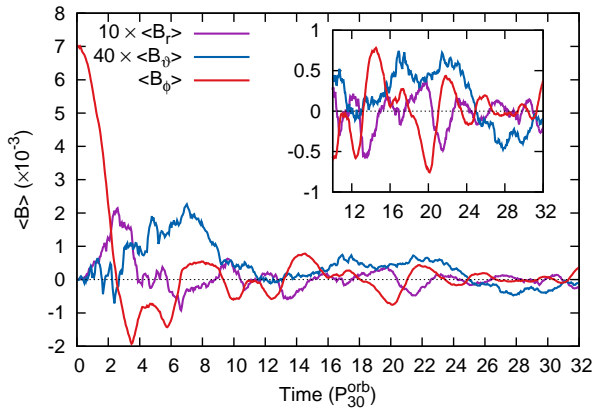}} \\
   \end{tabular}
   \caption{Volume-averaged magnetic field components for model
     gbl-sr. Note that $\langle B_{\rm r}\rangle$ and $\langle
     B_{\theta}\rangle$ have been multiplied by factors of 10 and 40,
     respectively, to aid comparison against the curve for $\langle
     B_{\phi}\rangle$.}
    \label{fig:meanB}
  \end{center}
\end{figure}

\subsection{The presence of a dynamo}

The time variability of the mean magnetic field components
(Fig.~\ref{fig:meanB}) is indicative of an $\alpha-\Omega$ dynamo in
our stratified global disk models. Furthermore, mean radial and
vertical fields develop within the first few orbital periods of the
simulation. The radial and azimuthal mean magnetic fields show
anti-correlated oscillations, the period of which is not obvious from
Fig.~\ref{fig:meanB}. This may be the result of averaging over a wide
range of radii \citep{O'Neill:2011}, or could be due to additional
terms contributing to the evolution of the mean fields when the
boundaries of the disk body are open - see the integrated induction
equations in \S~\ref{subsubsec:global}. A connection between the
vertical magnetic field and the radial and azimuthal fields is less
apparent, although there is a faint suggestion of oscillations in
$\langle B_{\theta}\rangle$ with a period on the order of $\sim
15~P^{\rm orb}_{30}$.

\section{Conclusions}
\label{sec:conclusions}

Global three-dimensional simulations of magnetorotationally turbulent
disks have been presented to investigate convergence with increasing
simulation resolution, magnetic energy, and quasi-steady
self-sustaining turbulence. A primary result of this work is
convergence with increasing resolution at an $\alpha$-parameter,
$\overline{\langle\alpha_{\rm P}\rangle}=0.04$, occurring at a
resolution of the order of 12-51 cells/$H$ in radius, 27 cells/$H$ in
the vertical direction, and 12.5 cells/$H$ in the azimuthal direction.
 
A control volume analysis applied to the body of the disk reveals the
dominant magnetic energy production to be the result of the
combination of Maxwell stresses and shear in the mean disk
rotation. Magnetic energy is primarily removed by dissipation, with a
negligible amount of energy being advected out of the disk body in
either the radial or vertical directions. Compressibility, or to be
more exact expansion, also contributes to the removal of magnetic
energy, but to a far lesser extent than dissipation. The control
volume analysis also allows the numerical resistivity of the
simulation code to be evaluated. The results reveal that sustained,
slowly diminishing turbulence can operate at $Re_{\rm M} \ltsimm
3000$, in contrast to the conclusions of \cite{Fleming:2000}, \cite
{Oishi:2011}, and \cite{Flock:2012b} that magnetorotational turbulence
should cease to function effectively at such values of $Re_{\rm
  M}$. This may be indicating that an effective large-scale dynamo can
operate at low magnetic Reynolds number in global disks.

The convergence with resolution found from our global simulations
occurs at roughly a factor of three lower resolution than found for
stratified shearing-box simulations by \cite{Davis:2010} \citep[see
also][and references there-in]{Shi:2010, Hawley:2011}. We have shown
how this result, as well as the convergence properties of unstratified
shearing-boxes \citep{Fromang:2007a, Simon:2009, Guan:2009} and global
disks \citep{Hawley:2011, Sorathia:2012} can be understood in terms of
balancing creation and dissipation of magnetic energy subject to
boundary conditions and magnetic field configuration. In particular,
using periodic boundary conditions in the radial direction (as in
shearing-box simulations) reduces the magnitude of a Lorentz force
term which depends on $B_{\rm r}$ and the radial gradient in
$B_{\phi}$. This term significantly contributes to magnetic energy
injection, and in global models (which use open radial boundaries) is
larger due to the presence of large scale radial gradients. Hence,
this term requires lower simulation resolution to achieve the same
power in a global model. Our results highlight important differences
between shearing-boxes and global disks which indicate the importance
of basing future deductions on stratified global models.

In closing we note that the results of this paper concern global disks
with a small net vertical magnetic field in the turbulent state. A
growing number of shearing-box studies are engaging in the challenging
task of modelling net flux magnetic fields in stratified disks
\citep{Suzuki:2009,
  Suzuki:2010, Moll:2012,
  Fromang:2013,Bai:2013,Lesur:2013}. Therefore, re-visiting the
analysis in this paper in the context of net vertical flux fields
would be a useful avenue for future work. Furthermore, the control
volume analysis we have used to derive the numerical resistivity could
be applied to recent orbital advection/FARGO schemes
\citep[e.g.][]{Johansen:2009, Stone:2010, Sorathia:2010, Mignone:2012}
to quantify their dissipation properties.

\subsection*{Acknowledgements}
We thank the anonymous referee for a useful report. This research was
supported under the Australian Research Council's Discovery Projects
funding scheme (project number DP1096417). E.~R.~P thanks the ARC for
funding through this project. This work was supported by the NCI
Facility at the ANU and by the iVEC facility at the Pawsey Centre,
Perth, WA.


 \newcommand{\noop}[1]{}

\appendix

\section{Fourier transform in spherical coordinates}
\label{sec:fts}

The simulations performed in this work use spherical polar
coordinates, so for consistency it is best to also perform the Fourier
transform in this coordinate system. We adopt spherical polars in real
space $(r,\theta,\phi)$ and in Fourier space $(k,\chi,\psi)$. That is,
\begin{eqnarray}
\begin{array}{r c l r c l r c l}
x & = & r \sin \theta \cos \phi, & k_x & = & k \sin \chi \cos \psi,\\
y &  = & r \sin \theta \sin \phi, & k_y & = & k \sin \chi \sin \psi,\\
z & = & r \cos \theta,  & k_z & = & k \cos \chi.
\end{array}
\end{eqnarray}

The following treatment is based on \citet{Baddour:2010} with minor
differences (primarily the notation of angles and the sign of $\bf k
\cdot x$ in the forward and inverse transforms).

\subsection{3D Fourier transform}

The Fourier transform of a function $f({\bf x}) = f(r,\theta, \phi)$ is

\begin{eqnarray}
F({\bf k}) = F(k, \chi, \psi) = \int_0^{2 \pi} \int_0^\pi \int_0^\infty f(r,\theta,\phi) \, e^{i {\bf k \cdot x}} \times\nonumber\\
r^2 \sin \theta \, dr \, d \theta \, d \phi.
\label{e:fft_3d}
\end{eqnarray}
Note that we use $e^{i \bf k \cdot x}$ here since this is consistent
with many definitions of the Fourier transform.

To proceed, both $e^{i {\bf k \cdot x}}$ and $f$ are expanded in terms
of spherical harmonics, which are defined by:
\begin{equation}
Y_l^m (\theta, \phi) = \sqrt{\frac{(2l+1)(l-m)!}{4 \pi(l+m)!}} \, P_l^m(\cos \theta) \,  e^{im\phi},
\end{equation}
where the $P_l^m(\cos \theta)$ are Legendre polynomials.

Let $j_l(z)$ be the spherical Bessel function of order $l$, then, denoting complex conjugates by *,
\begin{equation}
e^{i {\bf k \cdot x}} = 4 \pi \sum_{l=0}^\infty \sum_{m=-l}^{l} i^l \, j_l(kr) {Y_l^m}^* (\theta,\phi) 
Y_l^m(\chi,\psi), \label{e:exp_term}
\end{equation}
and,
\begin{equation}
f(r,\theta,\phi) = \sum_{l=0}^\infty \sum_{m=-l}^{l} f_l^m(r) Y_l^m (\theta,\phi), 
\end{equation}
where,
\begin{equation}
f_l^m(r) = \int_0^{2 \pi} \int_0^\pi f(r,\theta,\phi) \, {Y_l^m}^* (\theta, \phi) \> \sin \theta \, 
d\theta \, d \phi,
\label{eqn:flmr}
\end{equation}
are the spherical harmonic coefficients of $f(r,\theta,\phi)$.

With these expressions, the Fourier transform is:
{\setlength\arraycolsep{2pt}
\begin{eqnarray}
F(k,\chi,\psi) &=& 4 \pi \int_0^{2 \pi} \int_0^\pi  \int_0^\infty \left\lbrace \sum_{l=0}^\infty  \sum_{m=-l}^{l} f_l^m(r) 
Y_l^m (\theta,\phi) \right. \nonumber\\
&\qquad& \times \left.
\sum_{l^\prime=0}^\infty \sum_{m^\prime=-l^\prime}^{l^\prime} i^{l^\prime} j_{l^\prime}(kr) {Y_{l^\prime}^{m^\prime}}^*
(\theta,\phi) {Y_{l^\prime}^{m^\prime}}(\chi,\psi) \right\rbrace
\nonumber \\
 & \qquad & \times  r^2 \, \sin\theta \, d \theta \, d \phi. \label{eqn:full_FT}
\end{eqnarray}}

Using the orthogonality property of the spherical harmonics:
\begin{equation}
\int_0^{2 \pi} \int_0^\pi Y_l^m(\theta,\phi) \, {Y_{l^\prime}^{m^\prime}}^*(\theta,\phi) \, 
\sin \theta \, d \theta \, d \phi 
= \delta_{ll^\prime} \, \delta_{m m^\prime}, \label{eqn:sph_orth}
\end{equation}
we obtain,
\begin{eqnarray}
F(k, \chi, \psi) &=& 4 \pi \sum_{l=0}^\infty \sum_{m=-l}^l F_l^m (k) \, Y_l^m (\chi,\psi)  \label{eqn:F(k)} \\
\hbox{where,} \qquad F_l^m(k) &=& \int_0^\infty i^l r^2 j_l(kr) f_l^m(r) \> dr, \label{eqn:Flmk}
\end{eqnarray}
and the spherical harmonic coefficients $f_l^m(r)$ are given by
equation~(\ref{eqn:flmr}). The steps in evaluating the Fourier
transform are:
\begin{enumerate}[i)]
\item Evaluate Eq~(\ref{eqn:flmr}) for the spherical harmonic
  transform, $f(r,\theta,\phi) \Rightarrow f_{\rm l}^{\rm m}(r)$. \label{step:1}
\item Perform a spherical Bessel transform using Eq~(\ref{eqn:Flmk}),
  $f_{\rm l}^{\rm m}(r) \Rightarrow F_{\rm l}^{\rm m}(k)$.  \label{step:2}
\item The $F_l^m(k)$ are the complete set of Fourier coefficients and
  can be used to compute an angle averaged spectrum
  (\S~\ref{subsec:pow_spec}). One may perform an inverse spherical
  harmonic transform to acquire $F_{\rm l}^{\rm m}(k) \Rightarrow F(k,
  \chi, \psi)$ using Eq.~(\ref{eqn:F(k)}).
\end{enumerate}

For step~\ref{step:1} above we use the publicly available {\sc S2kit}
package\footnote{http://www.cs.dartmouth.edu/$\sim$geelong/sphere/} which
includes functions for performing spherical harmonic transforms on the
2-sphere using a combination of fast-Fourier transforms and
fast-Cosine transforms (to tackle the Legendre polynomials) and is
based on the seminal work by \cite{Driscoll:1994} \citep[see
also][]{Healy:2003}. The spherical Bessel transform is computed using
numerical quadrature in combination with a truncation of terms
contributing at large order $l$ to improve efficiency
(\S~\ref{subsec:j_l}).

\subsection{Angle-averaged spectrum} 
\label{subsec:pow_spec}
In the analysis of turbulence, one often uses the integrated energy spectrum:
\begin{equation}
 \Pi(k) = \int_0^{2 \pi} \int_0^\pi F({\bf k}) F^*({\bf k}) k^2 \> \sin \chi \, d\chi \, d \psi.
\end{equation}
Expressing $F({\bf k})$ in terms of the spherical harmonic expansion
(\ref{eqn:F(k)}), we have,
\begin{eqnarray}
\Pi(k) &=& \int_0^{2 \pi} \int_0^\pi \sum_{l=0}^B \sum_{|m|\leq l} 
\, \sum_{p=0}^B \sum_{|q| \leq p} F_l^m(k) {F_p^q}^*(k) \nonumber \\
&& \times Y_l^m(\chi,\psi) {Y_p^q}^*(\chi,\psi) \> \sin \chi \, d\chi \, d \psi \\
&=& \sum_{l=0}^B \sum_{|m|\leq l} F_l^m(k) {F_l^m}^*(k),
\end{eqnarray}
with the last equation resulting from the orthogonality of the
spherical harmonics (Eq~\ref{eqn:sph_orth}).

\subsection{Spherical Bessel functions for large $l$}
\label{subsec:j_l}

Eq~(\ref{eqn:Flmk}), which defines the $k$-dependence of the Fourier
coefficients, depends upon integration of the spherical harmonic
coefficients with the spherical Bessel functions $j_l(kr)$. These have
an interesting behaviour at large $l$; they are practically zero until
$kr \sim l$ following which they oscillate rapidly. The oscillatory
behaviour originates from the expression for the spherical Bessel
functions in terms of derivatives of the sinc function, \emph{viz.}
\begin{equation}
\label{bessel_sinc}
j_n(z) = (-1)^n z^n \left( \frac{1}{z} \frac{d}{dz}\right)^n \, \frac {\sin z}{z}
\end{equation}

That $j_l(kr) \approx 0$ for $kr \ll l$ follows from the leading term:
\begin{equation}
\label{e:lead_term}
j_n(z) =  \frac {2^n n!}{(2n+1)!} \,  z^n  + {\cal O} z^{n+1}
\end{equation} 
For what value of $z$ does $j_n(z)$ attain a numerically significant
value of, say, $\epsilon \sim 10^{-6}$? Take the logarithm of
equation~(\ref{e:lead_term}):
\begin{equation}
\label{log_jn}
\ln j_n(z) \approx n \ln 2 + n \ln z + \ln n! - \ln (2n+1)!
\end{equation}
and use Stirling's asymptotic form for the factorial function:
\begin{equation}
\label{e:stirling}
\ln n! \sim n \ln n - n
\end{equation}
to obtain
\begin{eqnarray}
\ln j_n(z) &\approx& n \ln 2 + n \ln z + n \ln n \nonumber \\
 & & - (2n+1) \ln (2n+1) + n + 1 \\
&=& \ln \epsilon \nonumber \\
\label{e:lnz}
\Rightarrow \ln z(\epsilon) &\approx &  \frac {1}{n} \ln \epsilon 
- \ln 2  +\frac{(2n+1)}{n} \ln (2n+1) \nonumber \\
& & - \ln n - \frac {n+1}{n} 
\end{eqnarray}
For example, for $\epsilon = 10^{-6}$ and $n=100$, $\ln z \approx
4.213$ and $z=67.57$.

\label{lastpage}

\end{document}